\documentclass[trackchanges,twocolumn]{aastex701}
\usepackage{siunitx}

\defcitealias{Meyer2013}{M+13}
\defcitealias{Snios2019}{S+19}

\begin{document}

\title{Resolving the temporal evolution of M87 jet with $\mathbf{\lesssim 0.1}$-arcsec \textit{Chandra} observations}

\author[orcid=0009-0001-5367-3976, gname=Camille, sname=Poitras]{Camille Poitras}
\affiliation{D\'{e}partement de Physique, de G\'{e}nie Physique et d’Optique, Universit\'{e} Laval, Qu\'{e}bec (QC), G1V 0A6, Canada}
\email[show]{camille.poitras.2@ulaval.ca}  

\author[orcid=0000-0002-4962-0740, gname=Gerrit, sname=Schellenberger]{Gerrit Schellenberger}
\affiliation{Center for Astrophysics | Harvard \& Smithsonian, 60 Garden Street, Cambridge, MA 02138, USA}
\email{gerrit.schellenberger@cfa.harvard.edu}

\author[orcid=0000-0002-0765-0511 , gname=Ralph, sname=Kraft]{Ralph Kraft}
\affiliation{Center for Astrophysics | Harvard \& Smithsonian, 60 Garden Street, Cambridge, MA 02138, USA}
\email{rkraft@cfa.harvard.edu}

\author[orcid=0000-0003-0297-4493, gname=Paul, sname=Nulsen]{Paul Nulsen}
\affiliation{Center for Astrophysics | Harvard \& Smithsonian, 60 Garden Street, Cambridge, MA 02138, USA}
\affiliation{ICRAR, University of Western Australia, 35 Stirling Hwy, Crawley, WA 6009, Australia}
\email{paulnulsen@gmail.com}

\author[orcid=0000-0002-7326-5793, gname=Marie-Lou, sname=Gendron-Marsolais]{Marie-Lou Gendron-Marsolais}
\affiliation{D\'{e}partement de Physique, de G\'{e}nie Physique et d’Optique, Universit\'{e} Laval, Qu\'{e}bec (QC), G1V 0A6, Canada}
\email{marie-lou.gendron-marsolais@phy.ulaval.ca}  

\author[orcid=0000-0002-7676-9962, gname=Eileen, sname=Meyer]{Eileen Meyer}
\affiliation{Department of Physics, University of Maryland Baltimore County, 1000 Hilltop Circle, Baltimore, MD 21250, USA}
\email{meyer@umbc.edu}

\begin{abstract}
    We present a 13-year \textit{Chandra}/HRC-I study of the M87 jet, using point-spread-function deconvolution to achieve unprecedented sub-arcsecond X-ray resolution of its kiloparsec-scale structure and temporal evolution. The deconvolved images reveal previously blended structures within the jet, including two components in HST-1 and complex morphology in downstream knots. Flux measurements reveal a global decrease in X-ray emission across the jet of up to $84\%$, consistent with synchrotron cooling. Modeling the fading yields minimum magnetic field strengths of $\sim324-1006~\mu$G for HST-1 and $\sim41-115~\mu$G for knot A. Consistent with synchrotron cooling, multi-wavelength comparisons with ALMA, JWST, and HST show that the principal X-ray structures closely match the jet width and knot locations observed at lower energies, while the X-ray emission is generally shifted upstream. Proper motion measurements show that jet features exhibit both quasi-stationary and superluminal apparent motions, reaching up to $4.8~c$ for HST-1, and also demonstrate that unresolved component blending can substantially bias inferred velocities. These results demonstrate the unique capability of high-resolution \textit{Chandra} X-ray imaging over long temporal baselines to resolve the evolving substructure of relativistic jets and to probe the particle acceleration and energy dissipation processes that shape their dynamics.
\end{abstract}

\keywords{galaxies: individual (M87) -- galaxies: active -- galaxies: jets -- X-rays: galaxies}


\section{Introduction} 

    Relativistic jets from active galactic nuclei (AGN) efficiently transport energy from accreting supermassive black holes (SMBHs) to galactic and intergalactic scales. However, the physical mechanisms responsible for energy dissipation and particle acceleration remain to be fully understood (e.g., \citealt{Sironi2015, Snios2019_Cen, Matthews2020}). Proposed mechanisms include shocks, interactions with obstacles, turbulence, velocity shear, and magnetic reconnection (e.g., \citealt{Blandford1978, BoschRamon2012, Stawarz2002, Nishikawa2005, Sironi2014}). These may coexist and operate across a wide range of spatial and temporal scales, making observations of jet morphology and variability essential to constraining particle acceleration models.

    X-ray observations provide one of the most direct probe of these processes. In large-scale Fanaroff-Riley type I (FR~I; \citealt{Fanaroff1974}) jets, the emission is widely interpreted as synchrotron radiation from electrons with Lorentz factors $\gamma \gtrsim 10^7$ (e.g., see \citealt{Harris2006} for a review). Their radiative lifetimes -- typically of a few to ten years -- are significantly shorter than the dynamical timescale of the flow. X-ray emission therefore traces sites of \textit{in situ} particle acceleration rather than advected electron populations. Observations of nearby jets reveal a combination of compact knots and diffuse emission (e.g. \citealt{Hardcastle2006, Sambruna2004, Goodger2010}), suggestive of multiple coexisting acceleration mechanisms. However, resolving these structures on sub-kiloparsec scales demands sub-arcsecond angular resolution, restricting detailed studies to only a handful of nearby systems, such as Centaurus A ($\sim3.8$~Mpc~; e.g., \citealt{Kraft2002, Harris2010, Snios2018, Bogensberger2024}).

    The jet in M87, at a distance of $16.5$~Mpc ($1\arcsec \approx 80~$pc; \citealt{Cantiello2024}), is highly suited for detailed study. Its proximity, combined with a $6.5\times10^9$~M$_\odot$ SMBH \citep{EHT2019}, makes M87 the only AGN from whose jet can be resolved from event-horizon scales to kiloparsec distances, while remaining bright across the entire electromagnetic spectrum, from radio to $\gamma$-rays (see \citealt{Hada2024} for a review). The jet extends over $\sim2$~kpc in a collimated, one-sided structure punctuated by a series of bright knots, including the prominent HST-1 complex ($\sim70$~pc from the core) which underwent a major multi-wavelength flare in 2005 \citep{Harris2006, Madrid2009}. High-resolution observations further reveal a complex, evolving transverse structure -- from spine-sheath configurations on parsec scales \citep{Walker2018, Cui2023} to a filamentary, double-helix morphology on kiloparsec scales \citep{Pasetto2021} -- suggesting a stratified flow and complex dynamics. Kinematic studies also reveal a gradual flow acceleration from sub-relativistic speeds near the core (e.g., radio: \citealt{Reid1989, Kovalev2007, Asada2014}) to superluminal motions downstream, reaching apparent velocities of up to $4-6c$ near HST-1 (e.g., radio: \citealt{Biretta1995}, optical: \citealt{Biretta1999, Meyer2013}, hereafter M+13, X-ray: \citealt{Snios2019}, hereafter S+19; \citealt{Thimmappa2024}). 

    A wealth of \textit{Chandra} studies has established that the M87 jet emits synchrotron X-rays and exhibits spectral variations, rapid variability (particularly in HST-1) and superluminal motions. Multi-wavelength comparisons reveal broadly similar structures across the spectrum, but with systematic spatial offsets between emission regions (e.g., \citealt{Marshall2002, Wilson2002, Harris2003, Perlman2005, Sun2018}; \citetalias{Snios2019}). Despite these advances, fully exploiting these observations requires X-ray imaging at spatial resolution comparable to that achieved at other wavelengths, together with long-term temporal baselines. The \textit{Chandra} High Resolution Camera (HRC), with the highest available angular resolution and negligible pile-up, provides a unique opportunity to address this limitation. In this work, we combine 13 years of HRC observations with advanced deconvolution techniques to obtain the sharpest X-ray view to date of the evolving M87 jet.
    
    This paper is structured as follows: Section~\ref{sec:observations} describes the observations and data reduction. The resolved jet substructure and its temporal evolution are investigated through flux variability (Section~\ref{sec:flux_variations_epochs}) and proper motion measurements (Section~\ref{sec:proper_motion}). Multi-wavelength comparisons are presented in Section~\ref{sec:multi-wavelength_comparison}, while the physical implications are discussed in Section~\ref{sec:discussions}, and conclusions are given in Section~\ref{sec:conclusions}. Throughout this paper, the nomenclature adopted for the knots follows that of \citet{Biretta1999}, \citet{Perlman1999}, \citetalias{Meyer2013}, and others.

\section{Observations} 
\label{sec:observations}
    To achieve the sub-arcsecond alignment accuracy required for our morphological and proper-motion study, we use \textit{Chandra} HRC-I observations, which are unaffected by photon pile-up and provide high angular resolution.
    We analyze a set of 14 observations of M87 listed in Table~\ref{tab:DetailsObservations}, spanning a temporal baseline of 13 years. These observations are grouped into four epochs -- 2012, 2017, 2023, and 2025 -- each with a comparable exposure of $\sim70$~ks. Two additional observations from 2019-2020 are available but are not included in the analysis since their combined exposure time (9.4~ks) is significantly lower.

    All observations were reprocessed using CIAO (version 4.18.0) with CALDB (version 4.12.3), via the standard \texttt{chandra\_repro} \citep{Fruscione2006}. After this reprocessing, a subset of the observations required special treatment. Specifically, the 2023 and 2025 observations contain corrupted secondary science data associated with unphysical instrument parameters. In these cases, all event data were retained, very short gaps in the good-time intervals were merged into a single continuous interval, and the deadtime correction factors were estimated by excluding anomalously low values and averaging the remaining measurements, all of which are close to unity.

    Background flares were removed from each dataset using CIAO's \texttt{deflare} tool with default parameters. The resulting clean exposure times are reported in Table~\ref{tab:DetailsObservations}. Exposure-corrected flux images were produced using \texttt{fluximage} for each observation, adopting the default HRC-I monochromatic energy of 1.5~keV, with a pixel size of $0.1318\arcsec$. The images were extracted within a rectangular region of $450\arcsec\times400\arcsec$ centered on the nucleus. The latest HRC-I background event file was automatically subtracted during flux-image creation.

    \subsection{Image alignment across epochs} 
            
            \begin{deluxetable*}{lcccc}
            \tablecaption{Summary of the HRC-I observations used in this analysis.
            \label{tab:DetailsObservations}}
            \tablewidth{0pt}
            \tabletypesize{\footnotesize}
            \tablecolumns{4}
                \tablehead{Obs. ID  & Exposure time\tablenotemark{a} [ks] & Date   & Shift\tablenotemark{b} (x, y) [pix] & Roll angle [$^\circ$]\tablenotemark{c}}
                \startdata
                \textit{2012 epoch} & & & & \\
                13515               & 74.31                           & 2012-04-14 & --             & 215.4  \\
                \hline
                \textit{2017 epoch} & & & & \\
                18612               & 72.48                           & 2017-03-02 & (-2.35, 0.40)  & 99.7 \\
                \hline
                \textit{2023 epoch} & \textit{70.92} & & & \\
                23685               & 14.07                           & 2023-07-03 & (7.32, -2.88)  & 248.2 \\
                27501               & 13.87                           & 2023-06-17 & (9.34, -1.83)  & 244.3 \\
                27502               & 14.26                           & 2023-04-18 & (6.38, -0.29)  & 211.9 \\
                27503               & 14.36                           & 2023-07-30 & (5.34, -3.87)  & 250.9 \\
                27504               & 14.36                           & 2023-05-14 & (5.02, -0.53)  & 245.9 \\
                \hline
                \textit{2025 epoch} & \textit{72.18} & & & \\
                30510               & 7.24                            & 2025-04-02 & (-4.38, -0.02) & 191.9 \\
                30546               & 9.48                            & 2025-03-16 & (-0.24, 2.27)  & 148.5 \\
                30547               & 9.93                            & 2025-03-21 & (-4.15, 0.75)  & 144.9 \\
                30548               & 10.72                           & 2025-06-15 & (-0.06, -4.07) & 234.9 \\
                30549               & 13.56                           & 2025-01-04 & (-0.53, 1.37)  & 54.9 \\
                30862               & 9.93                            & 2025-03-22 & (-1.95, 2.04)  & 144.9 \\
                30884               & 11.32                           & 2025-04-06 & (-7.53, 5.67)  & 186.9 \\
                \enddata
                \tablenotetext{a}{Exposure time after background flare removal.}
                \tablenotetext{b}{Relative shift applied with respect to the 2012 observation. Pixel scale: $0.1318\arcsec$ pix$^{-1}$.}
                \tablenotetext{c}{Nominal spacecraft roll angle (\texttt{ROLL\_NOM}).}
            \end{deluxetable*}
            
        To obtain reliable measurements of jet proper motions, all observations must be aligned to a common astrometric reference frame. For this purpose, the 2012 dataset (Obs.~ID 13515), which has the highest exposure time, was adopted as the reference image to which all other observations were aligned. An initial shift was estimated visually from the apparent core offset. The jet emission was masked to prevent it from dominating the alignment solution. Each image was also smoothed with a Gaussian ($\sigma = 2$~pix), which improves the robustness of the alignment procedure for observations with short exposure times. Using larger kernels yielded no significant improvement in the resulting astrometric accuracy.

        Because several short-exposure observations lack a sufficient number of well-defined point sources for reliable astrometric alignment, the image alignment was performed through a two-dimensional (2D) cross-correlation of the jet-masked flux images. A Lorentzian function was fitted to the cross-correlation function to determine the relative shifts between each observation and the reference frame. These shifts were then applied to the datasets using \texttt{wcs\_update}. Since the core remained close to the aim point in all observations (offsets of $\sim6''-30''$), geometric distortions are negligible. The 2023 and 2025 epochs are each composed of multiple short observations; following alignment, these data were combined using \texttt{merge\_obs}. The combined datasets have total exposure times of 70.92~ks and 72.18~ks, respectively, which are comparable to those of the 2012 (74.31~ks) and 2017 (72.48~ks) epochs.

        The alignment quality was assessed by measuring the centroid of the core in each epoch with \texttt{dmstat}. The alignment between the 2012 and 2017 epochs is accurate to within $0.004\arcsec$, while in the 2023 and 2025 epochs, the mean offsets of the individual observations relative to 2012 are below $0.012\arcsec$, with a standard deviation of $0.006\arcsec$. Although the limited counts in some of the shorter exposures naturally restrict the achievable precision, the overall alignment accuracy remains adequate for measuring jet proper motions. For context, a feature moving at the speed of light $c$ in M87 would shift by approximately $0.0039\arcsec$~yr$^{-1}$, which is larger than the misalignment found here.

    \subsection{PSF-deconvolved images} 
    \label{sec:psf-deconvolved_images}

        \begin{figure*}[!t]
            \centering
            \includegraphics[width=\textwidth]{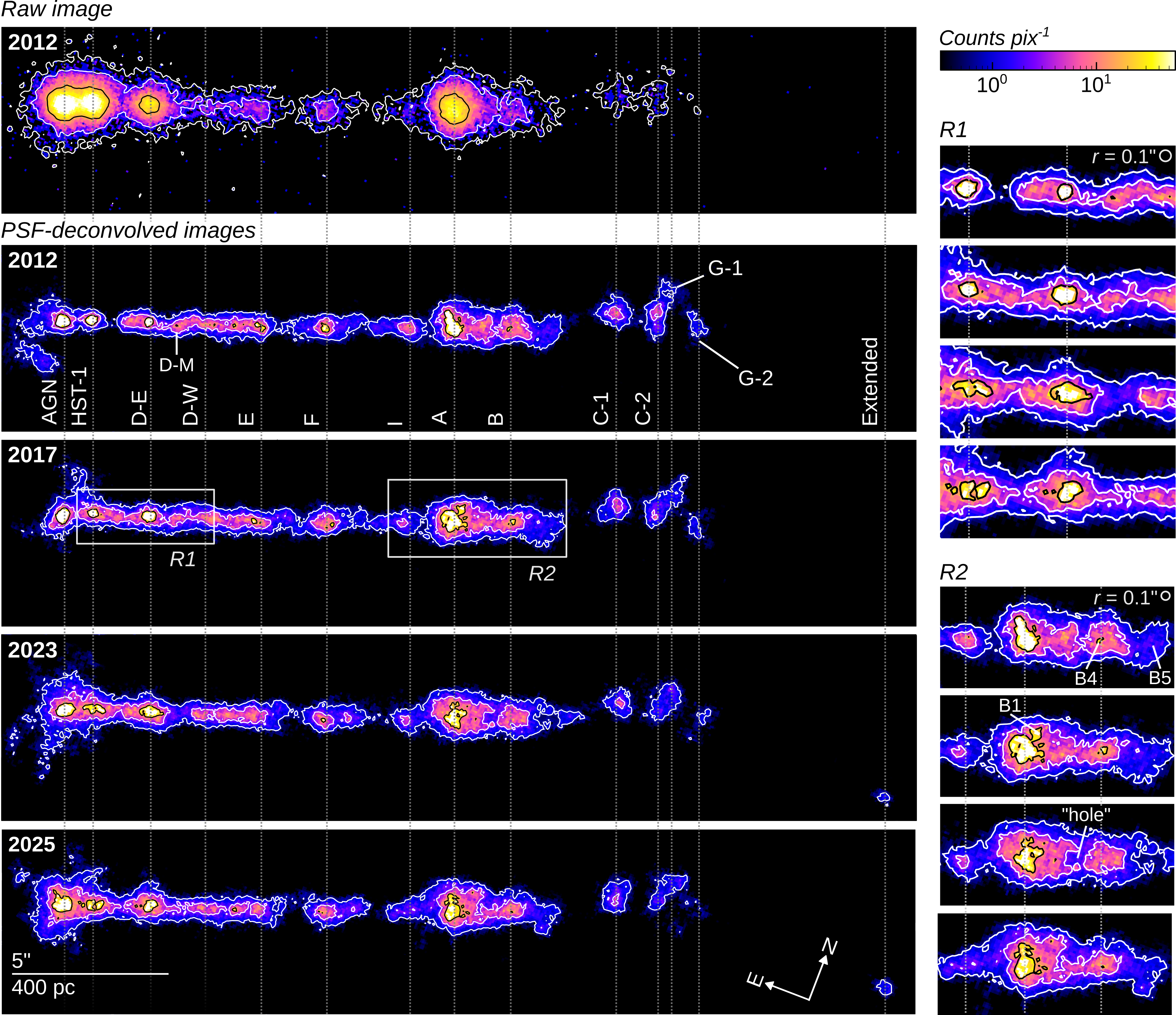}
            \caption{\textit{Left panel:} Raw (non-deconvolved) image from 2012 (top), followed by PSF-deconvolved images for each epoch (2012 to 2025, from top to bottom), shown at a pixel scale of $0.0659\arcsec$. \textit{Right panel:} zoomed-in regions for each epoch (same top-to-bottom order), corresponding to the areas indicated in the 2017 panel. The positions of the AGN and jet knots are marked by dotted vertical lines to guide the eye. All images are rotated by $21^\circ$ northward from west. Pixel-to-pixel intensities are not directly comparable between epochs.\label{fig:psf-deconv-emc2_epochs}}
        \end{figure*}
            
        To further investigate the sub-arcsecond morphology of the jet and fully exploit the sharp, well-sampled on-axis point-spread function (PSF) of HRC-I, we produced deconvolved images. Such analyses are commonly performed using ray-traced PSF simulations generated with \textit{Chandra} tools (e.g., \texttt{ChaRT}\footnote{\url{https://cxc.harvard.edu/ciao/PSFs/chart2/index.html}} and \texttt{MARX}\footnote{\url{https://chandra-marx.github.io/}}; \citealt{Wang2012, Posselt2012, Thimmappa2024}). However, our preliminary tests revealed a noticeable asymmetry in the reconstructed images, consistent with known discrepancies between simulated and observed \textit{Chandra} PSFs\footnote{\url{https://cxc.cfa.harvard.edu/ciao/caveats/psf_artifact.html}}. We therefore adopt empirical HRC-I PSFs derived from calibration observations of AR~Lac in the $\SIrange{0.8}{1.5}{keV}$ energy range\footnote{\url{https://cxc.cfa.harvard.edu/cal/Hrc/PSF/empPSF.html}}. Because the PSF evolves slowly with time due to thermal and instrumental effects \footnote{\url{https://iachec.org/wp-content/presentations/2025/mission_update_chandra.pdf}}, we use the 2006--2016 and 2017--2025 empirical PSFs for the corresponding epochs, with the characteristic PSF size increasing from $\sim0.37\arcsec$ to $\sim0.43\arcsec$ between the two periods. For each observation, the PSF was rotated to match the corresponding spacecraft roll angle (see Table~\ref{tab:DetailsObservations}). Both the count images and the associated PSFs were then rebinned to half the native HRC-I pixel scale ($0.0659\arcsec$~pix$^{-1}$), providing the finest sampling supported by our photon statistics. Because the source remained near the aim point in all observations, no additional corrections for off-axis PSF variations were required.

        Image deconvolution was performed using our implementation of the EMC$^2$ (\textit{Expectation through Markov Chain Monte Carlo}; e.g., \citealt{Esch2004, Karovska2005}) framework. In contrast to classical approaches (e.g., Richardson-Lucy; \citealt{Richardson1972, Lucy1974}), which primarily enhance compact features, EMC$^2$ provides a statistically robust framework for recovering both compact knots and diffuse structures simultaneously, making it well suited to the complex morphology of X-ray jets. 

        The deconvolution of the count images for each epoch was performed for 1500 iterations. For the 2023 and 2025 epochs, which consist of multiple observations, a single PSF was constructed by exposure-time-weighted averaging of the individual rotated PSFs of each ObsID. Deconvolution was then applied to the merged count image of each epoch to maximize photon statistics. We verified that performing the deconvolution on individual observations and subsequently combining the results yields consistent but slightly more blurred structures. The resulting deconvolved images are presented in Fig.~\ref{fig:psf-deconv-emc2_epochs}. Because the deconvolution redistributes counts among neighboring pixels while conserving the total number of counts, the resulting images are used solely for morphological analysis; pixel intensities are not directly comparable between epochs.
        
        Stability maps, defined as the ratio of the posterior mean to the posterior standard deviation of the EMC$^2$ image samples, were used to assess the robustness of the reconstructed features. Peak pixel values reach $\sim3-\SI{6}{\sigma~pix^{-1}}$ in the brightest knots, while fainter features are recovered at lower levels of $\sim0.5-\SI{1}{\sigma~pix^{-1}}$, compared with a background level of $\sim0.3-0.4~\sigma$~pix$^{-1}$. These values should be interpreted as measures of reconstruction stability rather than formal detection significances. Despite their lower stability levels, both the bright and faint structures are notably spatially consistent with emission observed at other wavelengths (see Section~\ref{sec:multi-wavelength_comparison}), supporting their physical origin.
        
        We further assessed the robustness of the deconvolution through two tests. First, repeating the deconvolution with PSFs rotated away from the observed spacecraft roll angle preserves the main jet features, but introduces low-level diffuse asymmetries with a common orientation around several knots, particularly around the brightest features. Second, we applied the same deconvolution procedure to the brightest nearby point source ($\lesssim 1\arcmin$ from the aim point) to assess the recovered angular resolution. In the deconvolved 2012 epoch, the source appears nearly point-like, with $\sim90$\% of the counts contained within the central $0.0659\arcsec$ pixel. For the 2017--2025 epochs, which use a different empirical PSF, the reconstructed emission remains confined within a radius of $0.1\arcsec$. The control source has a signal-to-noise (S/N) ratio of $\sim5-6$, comparable to that of most faint and diffuse jet structures. We also verified that the astrometric alignment is preserved in the deconvolved images, with AGN centroid offsets consistent with those measured in the aligned raw images. 

        The deconvolved images reveal rich substructure within the jet compared to the raw images. Several components exhibit clear morphological evolution between epochs, both along and across the jet, indicating complex dynamics on sub-arcsecond scales. We also identify a previously undetected feature emerging in the 2023--2025 data in the extended X-ray jet ($\sim26\arcsec$ from the core; see the region labeled "Extended" in Fig.~\ref{fig:psf-deconv-emc2_epochs}), which appears spatially offset from its counterpart at other wavelengths. Notably, the improved angular resolution achieved here enables a direct comparison with observations at other wavelengths, revealing closely matched structures (see Section~\ref{sec:multi-wavelength_comparison}). The global evolution of the jet and its implications for proper motions are further discussed in Section~\ref{sec:jet_temporal_evolution}.

\section{Flux Variations Between Epochs} 
\label{sec:flux_variations_epochs}
       \begin{figure*}[!t]
            \centering
            \includegraphics[width=\textwidth]{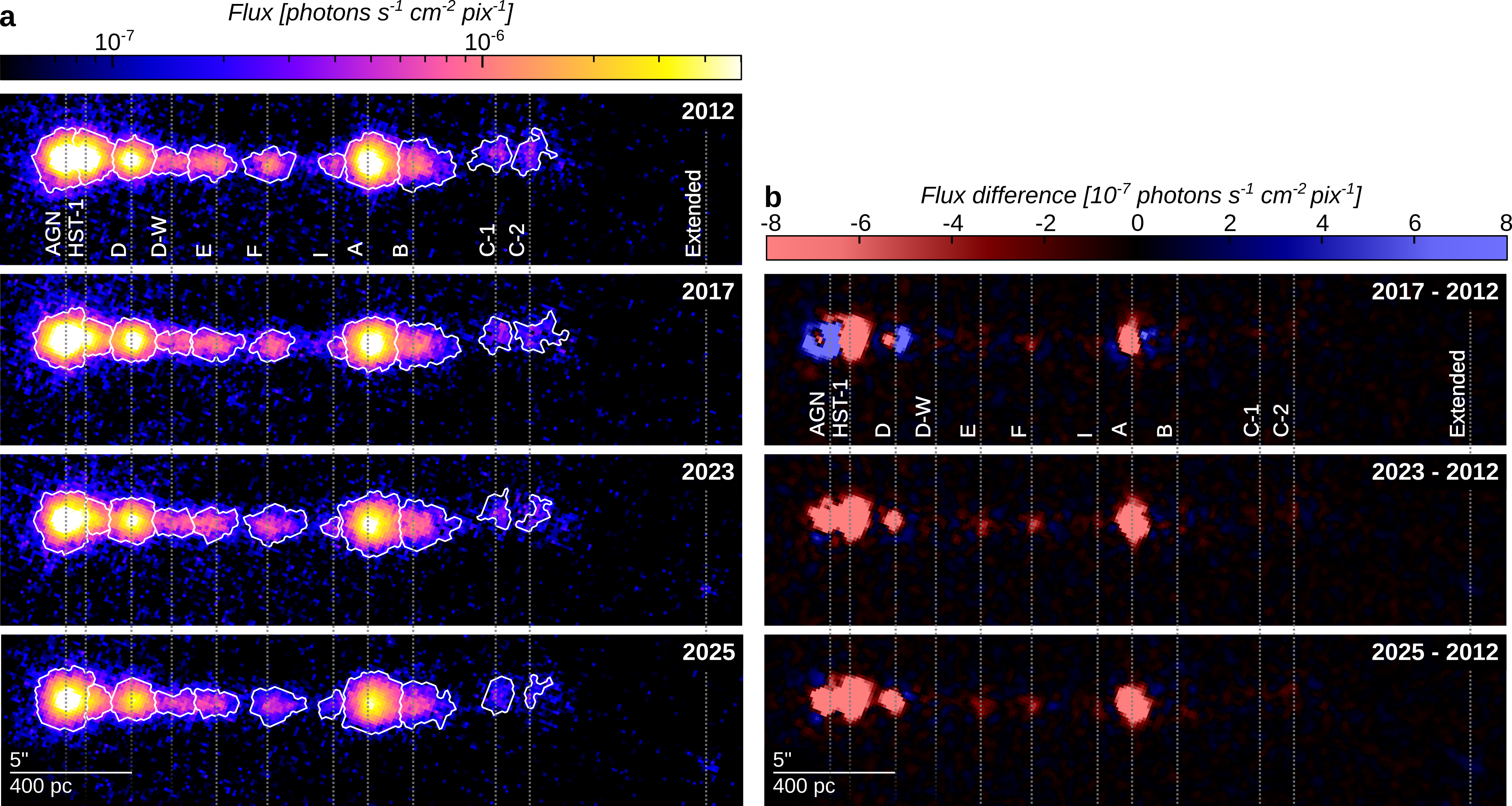}
            \caption{\footnotesize (a)~Aligned flux images for the 2012 to 2025 epochs (top to bottom), shown with a pixel scale of $0.1318\arcsec$~pix$^{-1}$, with the region used for the flux measurements of each jet feature overlaid as a solid white contour. (b)~Corresponding flux difference maps relative to the 2012 epoch, smoothed with a Gaussian kernel ($\sigma = 2$~pix). Blue/red regions indicate flux increases/decreases. For each panel, the AGN and jet knots are marked with dotted vertical lines.\label{fig:flux-diff_map}}
        \end{figure*}

        \begin{deluxetable*}{lcr@{${}\pm{}$}lr@{${}\pm{}$}lr@{${}\pm{}$}lr@{${}/{}$}lr@{${}/{}$}lr@{${}/{}$}l}
            \tablecaption{Flux measurements and relative flux variations of distinct features in the jet.\label{tab:flux_changes}}
            \tabletypesize{\footnotesize}
            \tablewidth{0pt}
            \tablecolumns{14}
                \tablehead{\colhead{Feature} & \colhead{Flux [$10^{-5}$ photon cm$^{-2}$ s$^{-1}$]} & \multicolumn{6}{c}{Change in flux\tablenotemark{a} [\%]} & \multicolumn{6}{c}{$R/N_P$ $\left[(S/N)^2/\mathrm{pixels}\right]$\tablenotemark{b}} \\
                \cline{2-2}\cline{3-8}\cline{9-14}
                \colhead{} & \colhead{2012} & \multicolumn{2}{c}{2017} & \multicolumn{2}{c}{2023} & \multicolumn{2}{c}{2025} & \multicolumn{2}{c}{2017} & \multicolumn{2}{c}{2023} & \multicolumn{2}{c}{2025}}
                \startdata
                AGN   & $87.8 \pm 4.7$ &  37 &  3 &  -4 &  2 &  -9 &  2 &  570.9 & 205 & 2880.1 & 205 & 3158.7 & 205 \\
                HST-1 & $70.9 \pm 3.9$ & -74 &  1 & -84 &  1 & -84 &  1 & 3217.4 & 160 & 5598.0 & 160 & 5537.8 & 160 \\
                D     & $30.0 \pm 1.7$ &  14 &  4 &  -2 &  3 & -18 &  3 &  206.3 & 136 &  864.9 & 136 &  985.5 & 136 \\
                D-W   & $ 5.1 \pm 0.5$ &  -7 & 10 &  13 & 10 &   4 & 11 &   91.9 &  76 &  168.6 &  76 &  137.6 &  76 \\
                E     & $ 8.2 \pm 0.6$ &  -9 &  7 & -25 &  6 & -34 &  6 &  151.6 & 122 &  340.8 & 122 &  291.4 & 122 \\
                F     & $ 6.4 \pm 0.5$ & -22 &  6 &  -1 &  7 & -11 &  7 &  141.1 & 120 &  296.0 & 120 &  261.3 & 120 \\
                I     & $ 2.2 \pm 0.3$ & -34 & 11 & -52 & 10 & -27 & 11 &   59.5 &  44 &   97.1 &  44 &   99.5 &  44 \\
                A     & $60.4 \pm 3.2$ & -10 &  2 & -23 &  2 & -27 &  2 &  554.1 & 232 & 2560.2 & 232 & 2490.4 & 232 \\
                B     & $10.7 \pm 0.7$ &   8 &  6 &  -4 &  6 &  -9 &  5 &  208.7 & 207 &  390.7 & 207 &  338.7 & 207 \\
                C-1   & $ 2.4 \pm 0.3$ & -14 & 10 & -26 & 10 &  -8 & 11 &   94.5 &  89 &  142.5 &  89 &  134.1 &  89 \\
                C-2   & $ 2.6 \pm 0.3$ &  -5 & 14 & -47 & 11 & -57 &  9 &   90.2 &  90 &  169.6 &  90 &  160.0 &  90 \\
                \enddata
                \tablenotetext{a}{Defined as $(F_{\mathrm{20XX}} - F_{\mathrm{2012}})/F_{\mathrm{2012}}$.}
                \tablenotetext{b}{$N_P$ corresponds to the number of pixels in the feature region as defined for the 2012 epoch.}
        \end{deluxetable*}
        
    Because the deconvolution redistributes counts among pixels, altering the local flux distribution while conserving the total number of counts, all flux variability analyses were performed on the raw flux images (pixel size of $0.1318\arcsec$~pix$^{-1}$). To enable a more accurate comparison of fluxes across epochs, the flux images were rescaled using the diffuse X-ray emission as an empirical reference. The local diffuse background, dominated by the hot gaseous component of M87, was measured in source-free regions at projected distances from the nucleus comparable to those of the jet. This approach is motivated by the expectation that the diffuse gaseous emission remains constant over our temporal baseline. We note that the non-X-ray background is $\sim30$ times lower than the diffuse emission and therefore does not significantly affect the rescaling. Each epoch was scaled to match the diffuse surface brightness measured in the 2012 reference observation. The levels measured in the 2017 and 2025 are consistent with those of 2012 (within 2\%, below the HRC systematic calibration uncertainties of $\lesssim5$\%), whereas the 2023 epoch exhibits a systematic excess of $\sim$10\%. Several point sources show similar small epoch-to-epoch variations, consistent with this trend. We verified that the derived scaling factors are insensitive to the precise choice of background region, within reasonable limits. Flux difference maps were subsequently produced by subtracting the 2012 reference image from the background-scaled flux images of each epoch using \texttt{dmimgcalc}. The flux and flux difference maps are shown in Fig.~\ref{fig:flux-diff_map}.

    To quantify the observed flux variations, we measured changes in the flux associated with individual features of the jet, including the AGN and individual knots. Regions associated with these features were defined using an image-based segmentation procedure applied independently at each epoch, allowing the region shapes to evolve with time. Starting from estimated feature locations, regions were grown to include contiguous emission exceeding the local background by a fixed threshold, while masking the neighboring features to avoid cross-contamination. This approach provides a more objective alternative to manual region selection and ensures a consistent definition of the regions across epochs. We verified that the inferred flux remains consistent within uncertainties under few-pixel shifts in the initial feature locations and variations in the background estimate and detection threshold. The resulting regions are shown in Fig.~\ref{fig:flux-diff_map}a.

    The total flux, inter-epoch flux differences, and associated uncertainties were measured for each region and epoch (see Table~\ref{tab:flux_changes}). Flux uncertainties include contributions from Poisson counting statistics, systematic calibration uncertainties ($\lesssim$~5~\% for HRC-I), and an additional component accounting for the sensitivity of the measurements to the precise definition of the region boundaries. The statistical significance of the inter-epoch flux variations was further evaluated using an integrated S/N metric similar to that used by \citet{Snios2018} and \citetalias{Snios2019}. For each pixel, the S/N was defined as
        \begin{equation}
            S/N = \frac{\left|c_2N_2 - c_1N_1\right|}{\sqrt{c_1^2N_1 + c_2^2N_2}} \;,
        \end{equation}
    where $N_1$ and $N_2$ are the raw counts measured at the two epochs, and $c_1$ and $c_2$ are the corresponding exposure-correction factors. To quantify variability associated with individual jet features, we computed the integrated squared S/N
        \begin{equation}
            R = \sum_i^{N_P}(S/N)_i^2 \;,
        \end{equation}
    where the sum is taken over the $N_P$ pixels comprising each feature region. For noise-dominated regions, $R$ is expected to follow a $\chi^2$ distribution with $N_P$ degrees of freedom, such that $R/N_P \approx 1$. As shown in Table~\ref{tab:flux_changes}, the AGN and several jet knots exceed the $3\sigma$ threshold. Statistically significant variability is detected in knots HST-1, D, and A during 2012–2017, and in all jet features at later epochs.

    The flux difference maps recover the behavior reported by \citetalias{Snios2019} for the 2012-2017 interval, in which HST-1 and knot A exhibit a net decrease in flux, while knot D shows evidence for apparent motion along the jet, characterized by a loss of flux upstream and a corresponding gain downstream. In contrast, the 2012-2023 and 2012-2025 comparisons are dominated by an overall decrease in flux across most of the jet. This behavior is consistent with synchrotron cooling and is discussed in more detail for the most prominent cases (i.e., HST-1 and knot A) in Section~\ref{sec:cooling_times_magnetic_field}.

\section{Proper motion} 
\label{sec:proper_motion}

    \subsection{Fractional difference maps} 
        As shown in Fig.~\ref{fig:flux-diff_map}b, the most recent epochs exhibit an overall decrease in flux relative to the 2012 observation across the jet. To isolate localized variations associated with motions and structural changes beyond the global flux evolution, we constructed fractional difference maps. Prior to differencing, the local background was subtracted. The jet was divided into four spatial subregions, separating brighter and fainter components, and each was normalized independently by its total flux. This approach, rather than normalization by the total jet emission, enhances localized variations in the fainter regions. We verified that the results are robust against reasonable changes in the normalization region or scaling factor: although the absolute amplitudes vary, the locations, signs, and relative trends that we aim to highlight remain unchanged.
        
\begin{figure}[!t]
                \centering
                \includegraphics[width=\columnwidth]{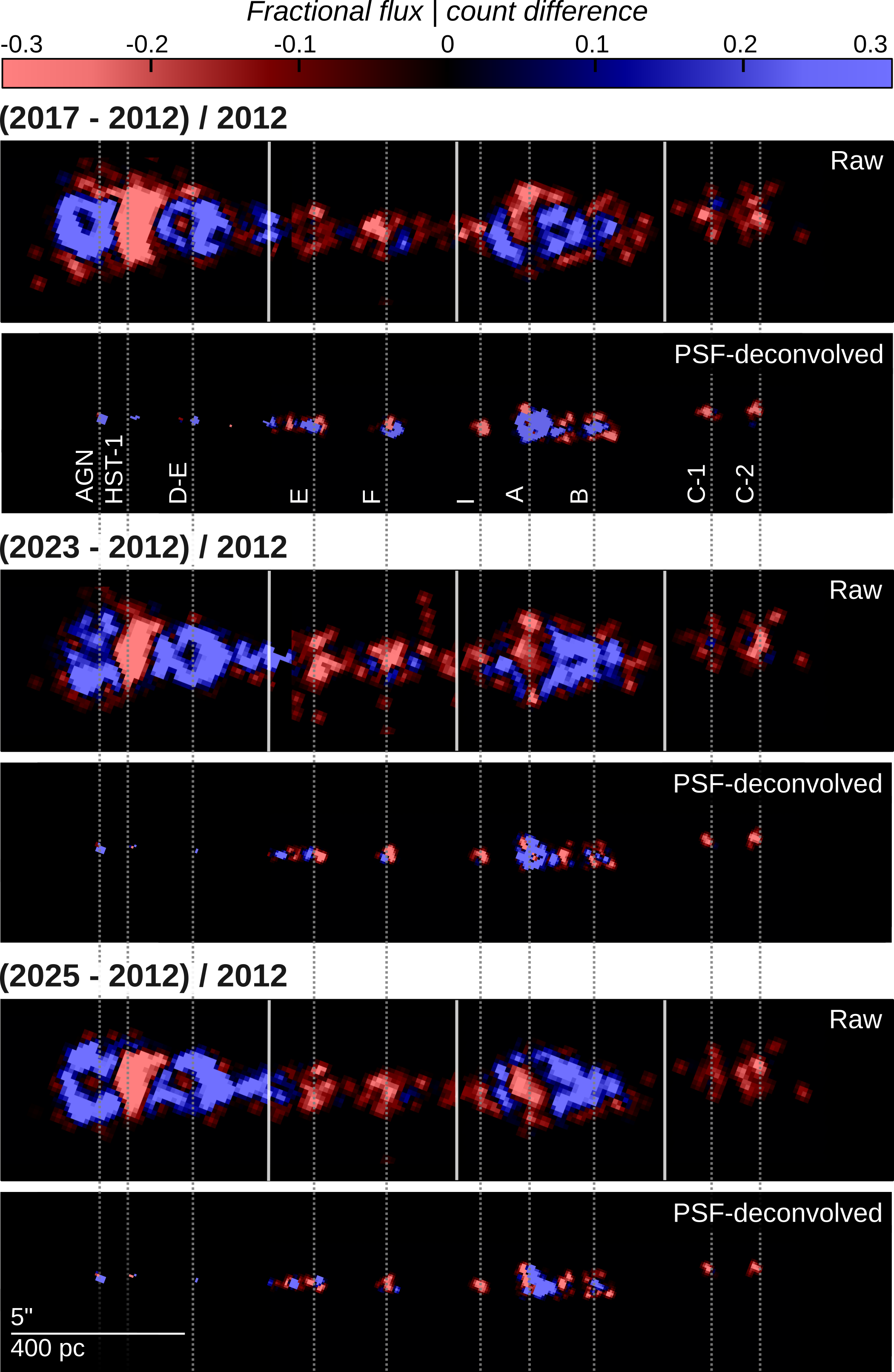}
                \caption{\footnotesize Fractional flux (raw, $0.1318\arcsec$~pix$^{-1}$) and count (PSF-deconvolved, $0.0659\arcsec$~pix$^{-1}$) difference maps relative to 2012 for each epoch (2017 to 2025, from top to bottom), defined as $(F_\text{20xx} - F_\text{2012})/F_\text{2012}$. The maps highlight spatial and morphological variations independent of absolute flux differences, with blue/red regions indicating flux increases/decreases. In the raw maps, the four jet subregions used for independent normalization are separated by solid vertical white lines. For each panel, the AGN and jet knots are marked with dotted vertical lines to guide the eye. All maps are smoothed with a Gaussian kernel ($\sigma =2$~pix).\label{fig:frac-diff-map}}
            \end{figure}

        The resulting fractional difference maps are shown in Fig.~\ref{fig:frac-diff-map}. For comparison, equivalent maps derived from the PSF-deconvolved count images, normalized to the total jet emission, are also presented. We note that the appearance of these maps depends on the chosen reference epoch. Fractional difference maps for all inter-epoch combinations are present in Fig.~\ref{fig:inter-epoch_frac-count-diff}.

        These maps reveal evolving substructures that cannot be explained by simple motion parallel to the jet axis alone, but instead suggest more complex behavior also including transverse displacements, which have also been observed in the optical with the \textit{Hubble Space Telescope} (HST) \citep{Meyer2013}. The PSF-deconvolved images further reveal more compact features. A quantitative analysis of the feature motions is presented in Section~\ref{sec:proper_motion_measurements}, while the physical implications of these results are discussed in Section~\ref{sec:jet_temporal_evolution}.

    \subsection{Proper motion measurements} 
    \label{sec:proper_motion_measurements}
            
            \begin{figure*}[!t]
                \centering
                \includegraphics[width=\textwidth]{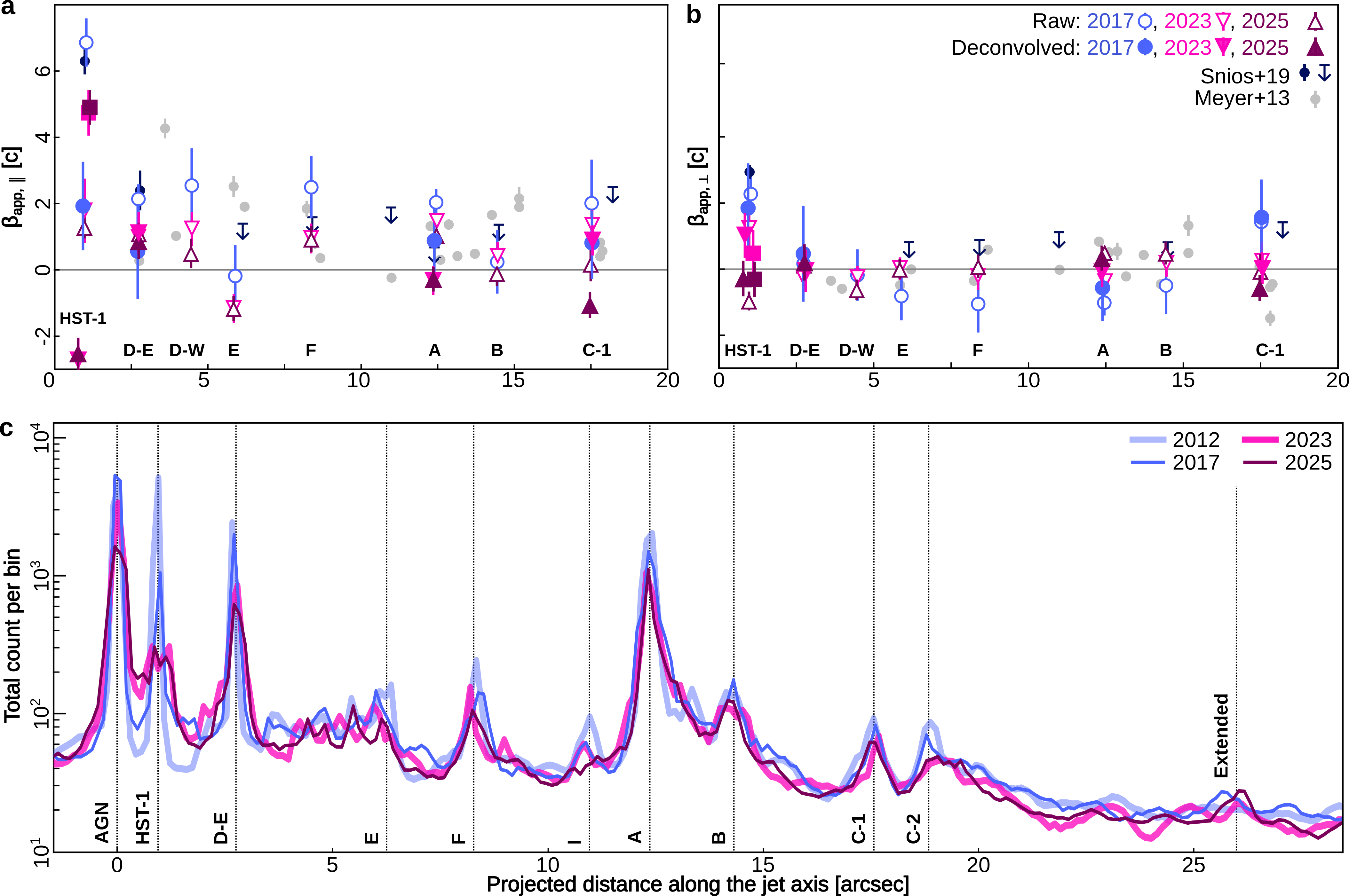}
                \caption{\footnotesize Apparent velocities of jet features relative to the 2012 reference epoch. Panels (a) and (b) show the velocity components parallel and perpendicular to the jet axis for 2017 (blue circles), 2023 (magenta downward triangles), and 2025 (purple upward triangles). Unfilled and filled markers indicate measurements from the raw and deconvolved images, respectively; square markers denote the HST-1-down component (2023, 2025). Previous optical \citepalias{Meyer2013} and X-ray \citepalias{Snios2019} measurements are shown for comparison. (c) Normalized count profiles along the jet axis from the deconvolved images for each epoch (2012: thick blue; 2017: thin blue; 2023: thick magenta; 2025: thin purple). Vertical dotted lines indicate the feature positions.\label{fig:proper-motion}}
            \end{figure*} 
    
        Proper motions are measured relative to the 2012 reference epoch using centroid positions measured for each jet feature. For a given feature, we used the regions defined in Section~\ref{sec:flux_variations_epochs}, masking all other emission. The centroid positions were computed using \texttt{dmstat}, which computes the count-weighted mean position. For the raw images, uncertainties are estimated from the standard deviation of the counts within the extraction region, scaled by the square root of the number of counts, i.e. $\sigma / \sqrt{N}$ (see \citetalias{Snios2019} for more details). All positions are measured relative to the AGN core. The projected displacements were then decomposed into components parallel and perpendicular to the jet axis ($21^\circ$ north of west), with positive perpendicular offsets defined upward in the rotated frame. From these offsets, we derive the apparent proper motion $\mu$, and the corresponding apparent velocity in units of the speed of light $\beta_\text{app}$.

        The centroid-based approach provides a consistent framework for measuring motions across the jet. However, because the AGN core and HST-1 are close in projection and partially blended in the raw images, centroid measurements alone are insufficient to determine their positions. We therefore performed 2D parametric modeling of this region (with a $0.0659\arcsec$~pix$^{-1}$ binning) using \texttt{Sherpa}, following the approach of \citetalias{Snios2019}. The emission is modeled as the sum of a constant background plane and two 2D Gaussian components representing the AGN and HST-1, with all parameters free to vary. Uncertainties on the fitted positions are estimated using the \texttt{conf} routine, which derives confidence intervals based on variations of the fit statistic. The resulting best-fit models and residual maps are presented in Fig.~\ref{fig:chandra-model_agn-hst1}. All the proper motions measured from the raw images are reported in Table~\ref{tab:proper_motions_raw}. 
        
        To mitigate the effects of component blending and exploit the enhanced angular resolution, we repeated the centroid-based proper-motion analysis using the PSF-deconvolved images. Because the deconvolution reveals a more complex morphology and, in some cases, multiple distinct components, cross-epoch identification is not always unambiguous. We therefore restrict this analysis to features that can be reliably tracked, namely the upstream and downstream components of HST-1 (hereafter HST-1-up and HST-1-down), and knots D-East (D-E), A and C-1. For these features, the centroid uncertainties are estimated from the residual spatial extent of the deconvolved structures, scaled by $1/\sqrt{N}$ to account for photon statistics. The resulting proper motions measured from the deconvolved images are listed in Table~\ref{tab:proper_motions_deconv}. A comparison with the corresponding raw-image measurements (Fig.~\ref{fig:proper-motion}) highlights the impact of component blending and suggests that the deconvolved measurements better trace the underlying motions. A discussion of the kinematic evolution of individual features and the effects of deconvolution on the inferred velocities is presented in Sections~\ref{sec:jet_temporal_evolution} and \ref{sec:resolution_blending_effects}.

            \setlength{\tabcolsep}{3.5pt}
            \begin{deluxetable*}{lcccccccccccccc}
            \tablecaption{Distances and proper motions measured from the deconvolved images for features tracked across epochs. \label{tab:proper_motions_deconv}}
            \tabletypesize{\scriptsize}
            \tablewidth{0pt}
            \tablecolumns{14}
                \tablehead{\colhead{Knot} & \colhead{Dist. [$\arcsec$]\tablenotemark{a}} & \multicolumn{3}{c}{$\mu_\parallel$ [mas yr$^{-1}$]} & \multicolumn{3}{c}{$\mu_\perp$ [mas yr$^{-1}$]} & \multicolumn{3}{c}{$\beta_{\mathrm{app},\parallel}$ [$c$]} & \multicolumn{3}{c}{$\beta_{\mathrm{app},\perp}$ [$c$]} \\
                \cline{3-5}\cline{6-8}\cline{9-11}\cline{12-14}
                \colhead{} & \colhead{} & \colhead{2017} & \colhead{2023} & \colhead{2025} & \colhead{2017} & \colhead{2023} & \colhead{2025} & \colhead{2017} & \colhead{2023} & \colhead{2025} & \colhead{2017} & \colhead{2023} & \colhead{2025}}
                \startdata
                HST-1\tablenotemark{b} & 0.91
                                 & 7.3(2.6)   & -       & - 
                                 & 6.8(2.6)   & -       & - 
                                 & 1.9(0.7)   & -       & - 
                                 & 1.8(0.7)   & -       & -  \\
                \;\textit{up}      & 0.80
                                 & -     & -10.2(1.3)   & -9.7(1.0)
                                 & -     & 3.6(1.3)     & -1.2(1.0)
                                 & -     & -2.7(0.7)    & -2.6(0.5)
                                 & -     & 0.9(0.7)     & -0.3(0.5) \\
                \;\textit{down}    & 1.11
                                 & -     & 17.9(1.3)    & 18.6(1.0)
                                 & -     & 1.6(1.3)     & -1.3(1.0)
                                 & -     & 4.7(0.7)     & 4.9(0.5)
                                 & -     & 0.4(0.7)     & -0.3(0.5) \\
                D-E              & 2.72
                                 & 2.1(2.5)   & 4.2(1.2)     & 3.3(1.6)
                                 & 1.6(2.5)   & -0.5(1.2)    & 0.6(1.6)
                                 & 0.6(0.7)   & 1.1(0.6)     & 0.9(0.5)
                                 & 0.4(0.7)   & -0.1(0.6)    & 0.2(0.5) \\
                A                & 12.38
                                 & 3.3(1.9)   & -1.2(0.8)    & -1.0(0.7)
                                 & -2.3(1.9)  & -0.6(0.8)    & 1.1(0.7)
                                 & 0.9(1.0)   & -0.3(0.4)    & -0.3(0.4)
                                 & -0.6(1.0)  & -0.2(0.4)    & 0.3(0.4) \\
                C-1              & 17.52
                                 & 2.9(1.9)   & 3.4(0.8)     & -4.0(0.7)
                                 & 5.7(1.9)   & -0.2(0.8)    & -2.3(0.7)
                                 & 0.8(1.0)   & 0.9(0.5)     & -1.1(0.4)
                                 & 1.5(1.0)   & -0.1(0.5)    & -0.6(0.4) \\
                \enddata
                \tablenotetext{a}{Projected distance from the core measured in the 2012 reference image (2023 for the resolved HST-1 substructures).}
                \tablenotetext{b}{The first row gives the single-component HST-1 measurement for 2012--2017; "up" and "down" denote the components resolved in 2023--2025.}
                \tablenotemark{\textbf{Notes.} Negative apparent motions likely result from unresolved substructures and their evolving flux contributions.}
            \end{deluxetable*}

\section{Multi-wavelength comparison} 
\label{sec:multi-wavelength_comparison}

        \begin{figure*}[!t]
            \centering
            \includegraphics[width=0.9\textwidth]{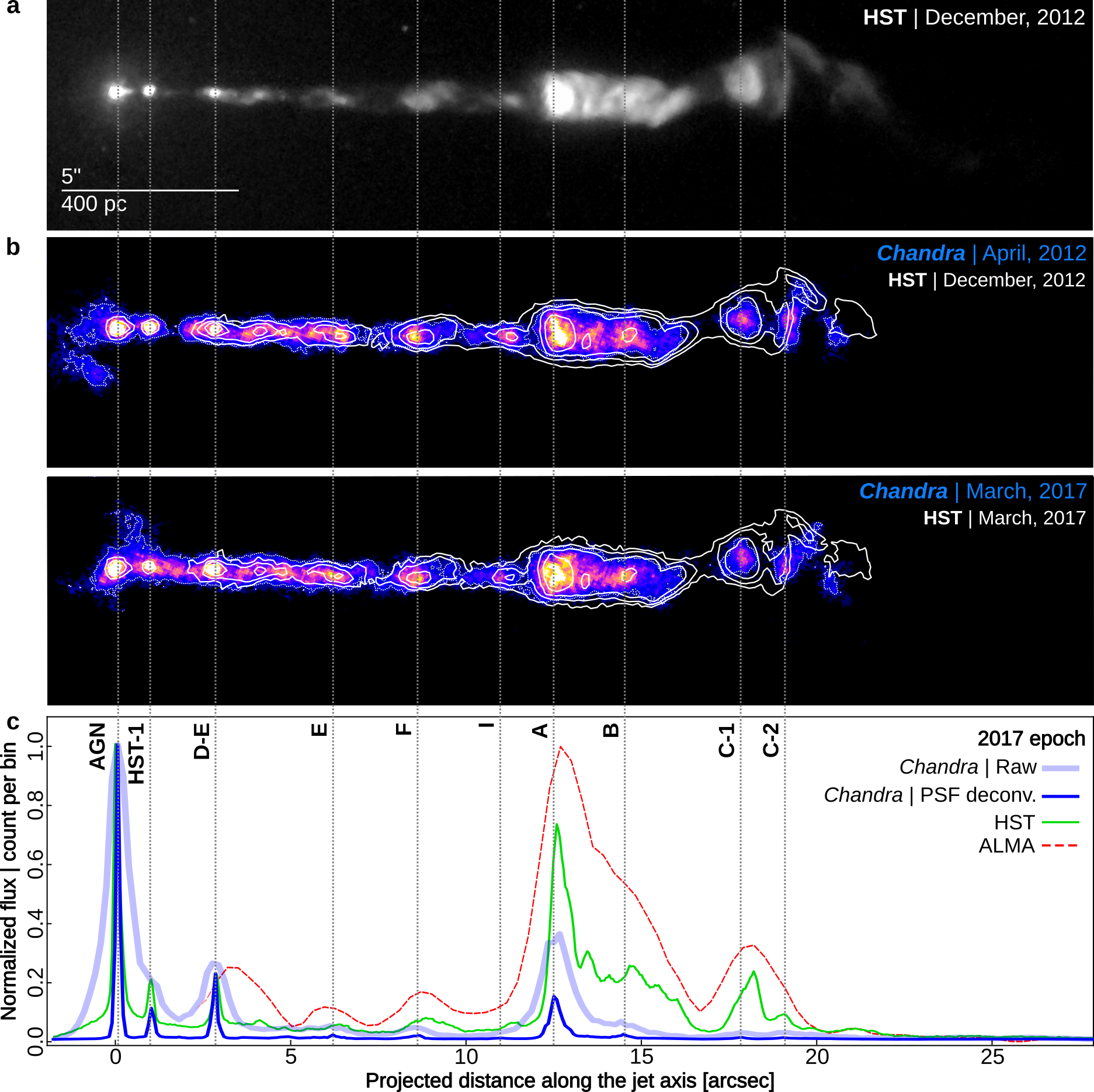}
            \caption{\footnotesize (a) HST/WFC3-UVIS image of the jet obtained in December 2012. (b) X-ray PSF-deconvolved images from April 2012 (top) and March 2017 (bottom), overlaid with HST contours from the corresponding epochs. (c) Radial profiles along the jet axis for 2017, showing the raw (thick light blue) and PSF-deconvolved (thin blue) X-ray data, the optical (green), and the ALMA data (dashed red; lower angular resolution). The strong unresolved emission from the AGN and HST-1 in the ALMA data is masked in the profile. Vertical dotted lines mark the AGN and knot positions.\label{fig:chandra-vs-hst}}
        \end{figure*} 

        \begin{figure*}[!t]
            \centering
            \includegraphics[width=0.9\textwidth]{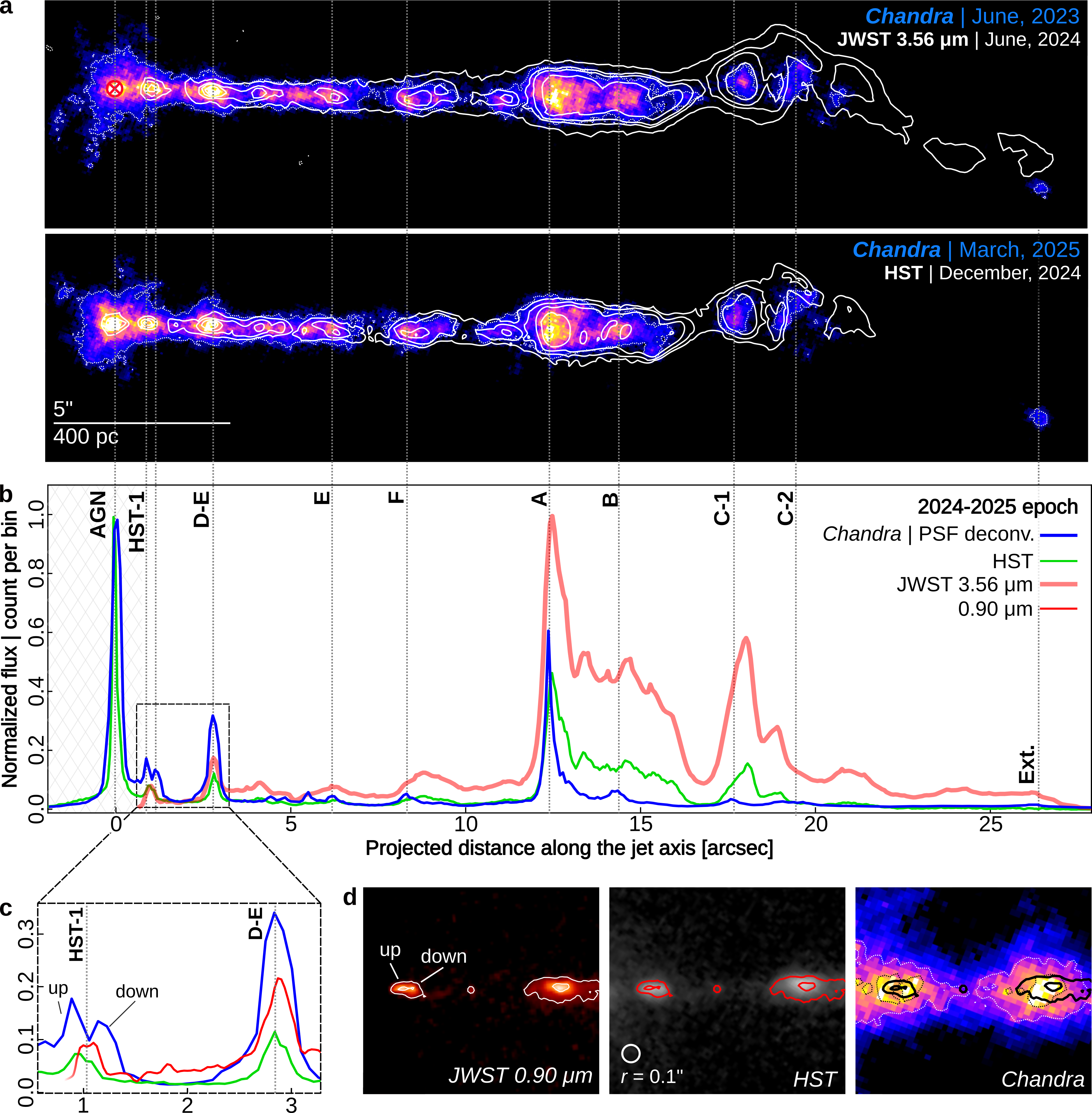}
            \caption{\footnotesize (a) PSF-deconvolved images of the jet for epochs centered on June 2023 (top) and March 2025 (bottom), overlaid with contours from JWST 3.56~$\mu$m (June 2024; top) and HST/WFC3-UVIS data (December 2024; bottom). The AGN position, masked in the JWST data, is marked by a red circle with a cross. (b) Radial profiles along the jet axis comparing the 2025 X-ray data (thin blue line) with the HST (green line) and JWST 3.56~$\mu$m (thick pink line). Vertical dotted lines mark the positions of AGN and knots. (c) Zoomed-in radial profiles for HST-1 and knot D, highlighting resolved substructures; the JWST profile corresponds to the $\SI{0.90}{\mu m}$ data. (d) Corresponding multi-wavelength images of the jet (2024--2025), with JWST contours (solid lines) overlaid in each panel.\label{fig:chandra-vs-hst-jwst}}
        \end{figure*} 

    Multi-wavelength comparisons provide an essential test of whether the knot morphology and motions observed in X-rays are consistent across bands. Because the synchrotron lifetimes of X-ray-emitting electrons are short, the X-ray knots are expected to lie closest to the sites of particle acceleration, making such comparisons a particularly valuable diagnostic. Early \textit{Chandra} observations established that the kpc-scale X-ray jet is broadly co-spatial with the optical and radio knots, while subsequent monitoring revealed correlated variability and superluminal motions, particularly in HST-1 (\citealt{Marshall2002, Harris2003}; \citetalias{Snios2019}). However, wavelength-dependent differences in peak locations and substructure have also been reported \citep{Perlman2005, Harris2009}.

    In this work, we revisit and extend these comparisons using our PSF-deconvolved X-ray images, which achieve angular resolution comparable to that of the optical, infrared and millimeter observations. The multi-wavelength analysis is performed by overlaying contours from the other wavelength data onto the X-ray images to assess structural correspondence and relative feature positions. Contour levels are selected to highlight both the global extent of the jet and the location of peak emission. To quantify potential offsets, we extract radial profiles along the jet axis within a narrow region closely tracing the jet emission. The binning matches the pixel scale of each instrument. The profiles are normalized to the AGN peak when present, and to knot A for data where the AGN is masked.

    We first compare the \textit{Chandra} observations with archival HST/WFC3-UVIS F275W images obtained in 2012 December 25 (ProgID 12989), 2017 March 3 (ProgID 14618), and 2024 December 11 (ProgID 17592). The WFC3/UVIS images have a pixel scale of $0.04\arcsec$~pix$^{-1}$. In addition, for the 2017 epoch we incorporate observations from the \textit{Atacama Large Millimeter Array} (ALMA) Band 6 (Project code 2016.1.01154.V; \citealt{Goddi2021}) obtained on 2017 April 10, covering the frequency range 211–275 GHz with a synthesized beam of $2\arcsec\times1\arcsec$. The results for these epochs are presented in Fig.~\ref{fig:chandra-vs-hst}. For the more recent epochs, we extend the comparison using infrared imaging obtained with the \textit{James Webb Space Telescope} (JWST; \citealt{Roder2025}), comparing the 2023 and 2025 \textit{Chandra} data with JWST observations acquired in June 2024 in the 0.90~$\mu$m ($0.031\arcsec$~pix$^{-1}$) and 3.56~$\mu$m ($0.063\arcsec$~pix$^{-1}$) bands. In both JWST images, the nucleus is saturated, and the 0.90~$\mu$m data additionally contain masked bad pixels in the extended jet region. The \textit{Chandra}-JWST comparisons are shown in Fig.~\ref{fig:chandra-vs-hst-jwst}.

    Across all wavelengths, the principal jet features are clearly detected and can be robustly cross-identified. While the overall morphology is broadly similar, systematic differences are apparent. In particular, the X-ray emission peaks are generally located slightly upstream of their optical and infrared counterparts, and variations in the relative brightness of individual knots are observed across wavelengths. Beyond knot C-2, the spatial correspondence between wavelengths becomes less uniform. A physical interpretation of these multi-wavelength trends is presented in Section~\ref{sec:multi-wavelength_synchrotron_cooling} and Section~\ref{sec:jet_temporal_evolution}.

\section{Discussion} 
\label{sec:discussions}

    \subsection{Synchrotron cooling} 
    \label{sec:synchrotron_cooling}

        \subsubsection{Cooling times and magnetic field estimates} 
        \label{sec:cooling_times_magnetic_field}
            To assess whether synchrotron cooling can account for the observed X-ray fading of the bright HST-1 and knot A features (see Section~\ref{sec:flux_variations_epochs}), we follow the formalism of \citetalias{Snios2019}. For synchrotron losses, the cooling time of electrons radiating at a fixed photon energy scales as $t'_{\rm syn} \propto B'^{-3/2}$, where primed quantities refer to the knot rest frame. To model the spectral evolution, we adopt the Kardashev–Pacholczyk (KP; \citealt{Kardashev1962, Pacholczyk1970}) framework, which assumes negligible pitch-angle scattering. The observed decrease in flux over a known time interval can therefore be used to constrain the minimum magnetic field strength required for synchrotron losses to reproduce the measured fading.
        
            For HST-1, the X-ray flux exhibits an exponential decline (see Fig.~\ref{fig:flux-evolution_HST1-A}), which we model as
                \begin{equation}
                    F(t) = C + (F_0 - C)\exp\left[-\frac{(t - t_0)}{\tau_{\rm obs}}\right],
                \end{equation}
            
            \noindent where $F_0$ is the flux at the initial epoch $t_0$, $C$ represents a constant baseline component, and $\tau_{\rm obs}$ is the characteristic decay time in the observer frame. The fit yields $\tau_{\rm obs} = (2.3 \pm 0.3)$~yr. In contrast, the flux of knot A decreases linearly over the monitoring period; we therefore adopt the observed interval between 2012 and 2025 as the characteristic fading timescale, $t_{\rm obs} = 12.93~{\rm yr}$.
            
            The intrinsic cooling time differs from the observed one because of relativistic effects. For a knot moving with bulk velocity $\beta c$ at angle $\theta$ to the line of sight, the intrinsic and observed timescales are related by
            \begin{equation}
                t' \approx \delta \, t_{\rm obs},
            \end{equation}
            
            \noindent which is appropriate for the small redshift of M87. Following \citetalias{Snios2019}, we estimate the beaming factor under the minimum-momentum assumption as $\delta_{\rm m} = \sqrt{1 + \beta_{\rm app}^{2}}$. Because the apparent velocity varies over time (see Section~\ref{sec:proper_motion}), we adopt the minimum and maximum measured values of $\beta_{\rm app}$ to bracket the allowed range of intrinsic cooling times. The intrinsic decay times are $4.0-15.0$~yr for HST-1 and $18.3-30.1$~yr for knot A. 
            
            Using these intrinsic timescales and the flux decrease measured from the raw (non-deconvolved) images (see Table~\ref{tab:flux_changes}), we estimate the minimum magnetic field strength required for synchrotron cooling to reproduce the observed fading between 2012 and 2025. Following \citetalias{Snios2019} and the KP model, we assume an initial power-law electron distribution $dN/d\gamma = K\gamma^{-p}$, neglect pitch-angle scattering (thereby maximizing the cooling rate), and evolve the distribution under synchrotron losses over the intrinsic time interval. We adopt $p = 3.6$ for HST-1 and $p = 4.2$ for knot A, following the spectral modeling of \citet{Kataoka2005}. The magnetic field $B'$ is adjusted until the modeled X-ray flux reduction matches the observed value. Under these assumptions, we obtain minimum magnetic field strengths between $324$ and $1006~\mu$G for HST-1 and between $41$ and $115~\mu$G for knot A. These values are broadly consistent with previous estimates obtained using independent techniques, though they should be interpreted as lower limits rather than direct measurements of the total field strength. For HST-1, equipartition and variability-based analyses suggest magnetic field strengths of order $\sim600$--$1000~\mu$G \citep{Harris2003, Harris2009}, while the cooling-based estimate of \citetalias{Snios2019} is $B' \sim 420~\mu$G. For knot A, equipartition estimates of $B \sim 330~\mu$G \citep{Kataoka2005} and the cooling-based estimate of $B' \sim 230~\mu$G \citepalias{Snios2019} are slightly higher than the minimum field strengths inferred here.

        \subsubsection{Multi-wavelength signatures of synchrotron cooling} 
        \label{sec:multi-wavelength_synchrotron_cooling}
            Additional evidence for synchrotron cooling is provided by the multi-wavelength morphology of the jet. As shown in Section~\ref{sec:multi-wavelength_comparison}, the main jet features are detected across millimeter, infrared, optical, and X-ray. Despite this overall correspondence, systematic differences are observed in the relative positions and brightness of the emission peaks. In particular, the X-ray emission is generally located slightly upstream of its multi-wavelength counterpart. This is evident in knots D-E, E, and F, where the X-ray peaks are offset by $\sim0.07\arcsec$, $\sim0.2\arcsec$, and $\sim0.5\arcsec$, respectively, relative to the optical emission (measured in the 2024--2025 epoch). Similar behavior is observed in knot A, where the X-ray emission is offset by $\sim0.14\arcsec$ and traces a sharper, more upstream structure than at other wavelengths. These results are broadly consistent with previous measurements, which typically find X-ray-to-optical offsets of $\sim0.05\arcsec-0.2\arcsec$ (e.g., \citealt{Marshall2002, Perlman2005}). Such offsets are naturally expected in a synchrotron cooling scenario, where higher-energy electrons responsible for the X-ray emission have shorter radiative lifetimes and therefore emit closer to their acceleration sites, whereas lower-energy electrons can propagate farther downstream before fading.
            
            This energy-dependent behavior is also reflected in the relative flux distribution of individual knots. Variations are observed in the brightness contrast between knots across wavelengths. In particular, knot B appears comparatively brighter than knot A at lower energies, while this contrast is reduced in X-rays, as quantified by the flux (or count) ratios $F_A/F_B \approx 8$ (X-ray), $\sim2.5$ (optical), $\sim2$ (infrared), and $\sim1.7$ (millimeter). A similar trend is observed between knot D-E and downstream features. These variations indicate that lower-energy emission traces particles advected downstream over longer distances, as their synchrotron cooling times are orders of magnitude longer than those of the X-ray emitting electrons.
        
            However, beyond knot C-1, the multi-wavelength behavior becomes less consistent with the simple synchrotron cooling picture. In this extended region, the X-ray emission can be offset differently, in some cases appearing downstream, and fading less rapidly than at longer wavelengths. This suggests that additional processes are at play (see Section~\ref{sec:downstream_knot-C} for further discussion).

    \subsection{Jet temporal evolution} 
    \label{sec:jet_temporal_evolution}
\begin{figure*}[!t]
                \centering
                \includegraphics[width=\textwidth]{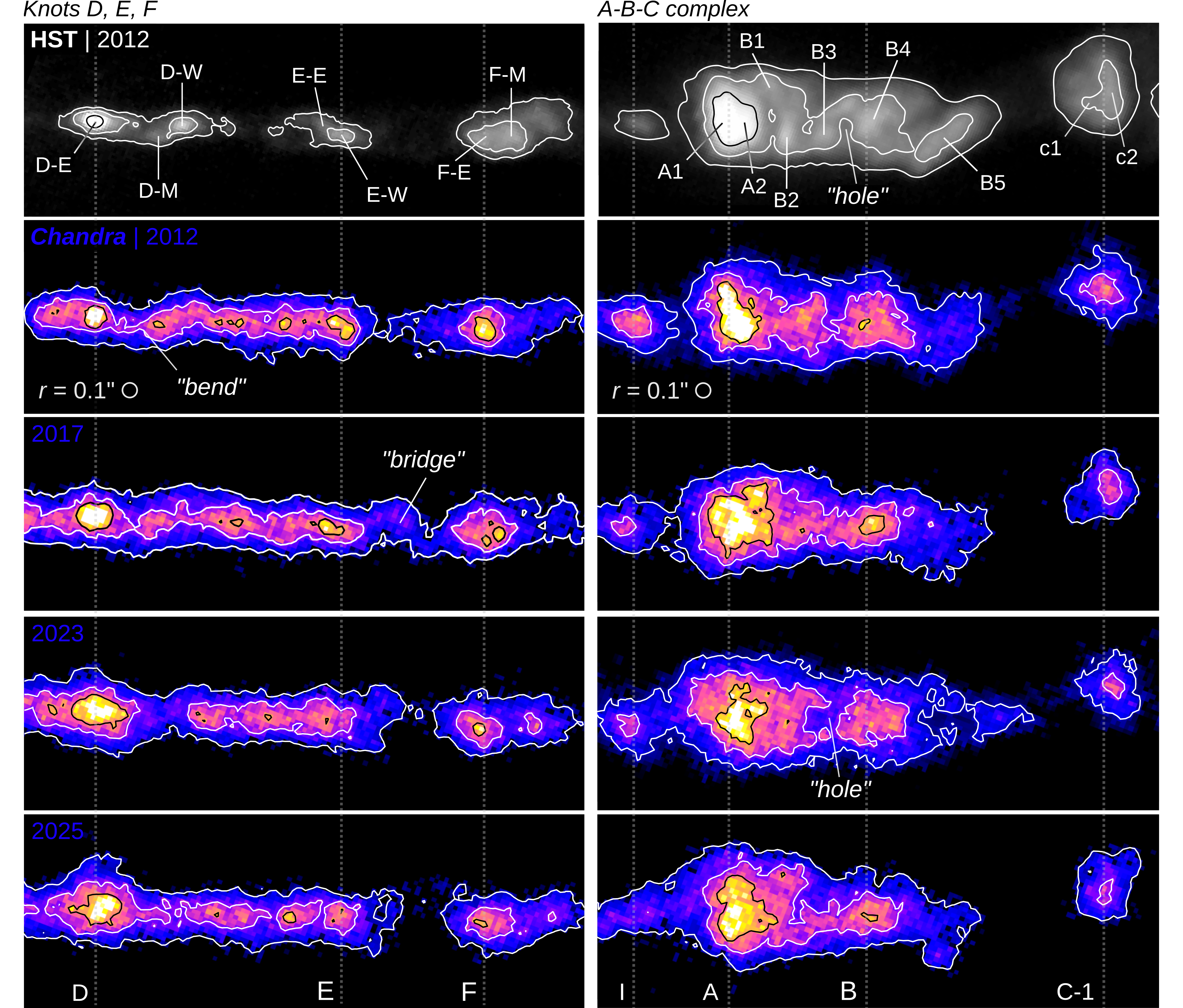}
                \caption{Comparison between HST and PSF-deconvolved \textit{Chandra} images of (\textit{left}) knots D--F and (\textit{right}) the A--C complex. The top panels show the 2012 HST image with the subcomponent nomenclature adopted from \citealt{Biretta1999} and \citetalias{Meyer2013}. The lower panels show the \textit{Chandra} images from 2012 to 2025, from top to bottom. Vertical lines indicate the positions of the principal knots.\label{fig:subcomponents_hst-chandra}}
            \end{figure*}

        The temporal evolution of the jet between 2012 and 2025 reveals significant changes in both the longitudinal and transverse directions. The deconvolved images uncover a rich substructure within the X-ray jet, enabling us to track apparent motions and morphological evolution on sub-arcsecond scales. In the following, we describe the behavior of individual features from the core to the outer jet, combining X-ray measurements from the deconvolved images with comparisons at other wavelengths. The knot nomenclature used throughout this section is illustrated in Fig.~\ref{fig:psf-deconv-emc2_epochs}, while Fig.~\ref{fig:subcomponents_hst-chandra} summarizes the subcomponent nomenclature for the D--F (Section~\ref{sec:knots_D-E-F-I}) and A--C (Section~\ref{sec:A-B-C_complex}) complexes based on previous optical studies. Unless otherwise stated, all distances are given relative to the core. The impact of deconvolution and unresolved components is further discussed in Section~\ref{sec:resolution_blending_effects}.
        
        \subsubsection{HST-1} 
            HST-1 is the most prominent feature in the jet and has received considerable attention in previous studies due to its complex behavior and variability (e.g., \citealt{Harris2006, Cheung2007, Hada2024, Madrid2009, Thimmappa2024}). In the raw X-ray images, HST-1 appears as a single unresolved feature across all epochs. However, the enhanced angular resolution provided by the deconvolved images resolves two distinct components from 2023 onward, which closely correspond to and align with those observed in optical and infrared wavelengths.

            The improved imaging has a direct impact on the inferred kinematics. In the raw images, the close projected separation of the AGN and HST-1 makes it difficult to disentangle their emission, so we use 2D modeling to estimate their positions, yielding $\beta_{\text{app}, \parallel} = (6.9 \pm 0.6)c$ and $\beta_{\text{app}, \perp} = (2.2 \pm 0.5)c$ between 2012 and 2017. These values are consistent with previous X-ray analyses of the same dataset \citepalias{Snios2019} and with long-monitoring \citep{Thimmappa2024}. However, the deconvolved images clearly separate the AGN and HST-1 (located at $0.91\arcsec-0.95\arcsec$), enabling more robust centroid measurements. Over the same interval, the inferred parallel velocity is significantly lower, with $\beta_{\text{app}, \parallel} = (1.9 \pm 0.7)c$, in better agreement with optical measurements of $\sim2.1c$ over 2005-2022 \citep{Thimmappa2026}. 

            In the later epochs, the raw images already show a decrease in apparent velocity, with $\beta_{\text{app}, \parallel} = (1.8 \pm 0.9)c$ in 2023 and $(1.4 \pm 0.3)c$ in 2025. This rapid change relative to 2017 suggests that the emission is not dominated by a single component; when multiple components are unresolved, the measured centroid traces a flux-weighted average of their positions, yielding an apparent velocity that can vary. The deconvolved images confirm this by resolving two distinct components within HST-1, located at $0.80\arcsec$ and $1.11\arcsec$ from the core in 2023. Relative to the 2012 epoch, the upstream and downstream components exhibit persistent superluminal motion $\beta_{\text{app}, \parallel} = (-2.7 \pm 0.6)c$ and $(4.7 \pm 0.6)c$, respectively. The negative apparent velocity of the upstream component does not imply physical motion backwards, but most likely reflects its position relative to the earlier unresolved centroid, which was likely the flux-weighted combination of these two components. Back-extrapolating the 2023-2025 component motions (assuming constant velocities) to 2017 places the two subcomponents at projected distances of $\sim0.07$ and $\sim0.08$~kpc, i.e., separated by only $0.01$~kpc ($0.08\arcsec$), and thus unresolved. For equal flux contributions, the corresponding blended centroid would imply an apparent velocity of order $\sim1c$, which is closer within uncertainties to the deconvolved 2012-2017 measurement of $(1.9 \pm 0.7)c$. The measured centroid velocity would, however, depend on the relative flux of the two components, which evolves with time as shown in Fig.~\ref{fig:proper-motion}c.

            Long-term VLBI monitoring (2004--2020) of HST-1 shows quasi-periodic ejection of multiple superluminal subcomponents from a more stationary upstream edge of the complex, often referred to as HST-1d (e.g., \citealt{Cheung2007, Giroletti2012, Hada2024}, see their Fig.~12). These components exhibit apparent velocities in the range of $\sim4.5-5.5c$. Assuming roughly constant velocities, a simple extrapolation suggests that some of these components would reach projected distances of $\sim1.11-1.18\arcsec$ from the core by 2025, placing them close to the downstream X-ray component of HST-1 ($\sim1.15\arcsec$). Given their small separations and the angular resolution of our X-ray data, such components would likely be unresolved and appear as a single downstream feature. In this context, the upstream X-ray emission near the nominal HST-1 position may trace the quasi-stationary region (HST-1d) together with newly ejected components that have not yet propagated significantly downstream. The downstream X-ray component also exhibits a comparable velocity ($\sim4.7c$), broadly consistent with the radio measurements; however, any direct association should be treated with caution given the extrapolated nature of the comparison and the known structural evolution of HST-1.

        \subsubsection{Knots D, E, F, and I} 
        \label{sec:knots_D-E-F-I}
        
            At knot D, the emission is dominated by the D-E component ($\sim2.7\arcsec$ from the core), for which we measure $\beta_{\text{app}, \parallel} \approx (0.87 \pm 0.35)c$ and $\beta_{\text{app}, \perp} \approx (0.17 \pm 0.35)c$ from the deconvolved images, consistent with a relatively stationary feature. Morphologically, the emission evolves from a roughly circular to elongated structure, suggesting the emergence of closely spaced subcomponents (separated by $\sim0.2\arcsec$). These results are broadly consistent with optical measurements, which report a lower velocity of $(0.28 \pm 0.05)c$ for D-E and interpret it as a quasi-stationary feature through which faster components may propagate \citepalias{Meyer2013}. Knot D is also known to host at least two additional components, D-Middle (D-M) and D-West (D-W), with optical measurements indicating parallel motions of approximately $4.3c$ and $1c$, respectively \citepalias{Meyer2013}. Polarimetric studies further reveal additional substructures with distinct magnetic field configurations (e.g., \citealt{Perlman1999, Avachat2016}). In our X-ray data, the regions associated with D-M and D-W (between $\sim3\arcsec-4.2\arcsec$) exhibit pronounced variability in the radial profiles (see Fig.~\ref{fig:proper-motion}c), along with clear transverse evolution. In particular, the jet shows a localized bend near D-M in 2012 -- visible as a downward deviation between D-E and D-W (see Fig~\ref{fig:subcomponents_hst-chandra}) -- that becomes more aligned with the jet axis by 2025. However, these components cannot be reliably tracked across the $\sim 13$~yr baseline, nor even between the two most recent epochs. This behavior suggests that D-M and D-W are short-lived, rapidly evolving structures.

            Similar complex behavior is observed in knots E and F. At knot E ($\sim5.9\arcsec$), the X-ray emission shows evidence for substructure, with two components separated by $\sim0.2\arcsec$ in 2012. At later epochs (2023--2025), these components appear blended. At knot F ($\sim8.4\arcsec$), the subcomponents identified in optical studies (e.g., F-E moving at $\sim1.9c$ and F-M at $\sim0.4c$; \citetalias{Meyer2013}) are not resolved in our X-ray data. Nevertheless, the raw-image proper motions suggest a transition in the dominant emitting component, with the apparent parallel velocity decreasing from $(2.5 \pm 0.9)c$ in 2017 to $\sim(0.9 \pm 0.5)c$ in the later epochs. The faint and more diffuse transverse structure between knots E and F (labeled "bridge" in Fig.~\ref{fig:subcomponents_hst-chandra}) also evolves with time, with the diffuse emission changing from a locally inclined morphology, nearly perpendicular to the jet axis in 2017, to a more aligned configuration in 2025. This could indicate a local oscillation or helical perturbation in the flow.

            A feature associated with knot I ($\sim11\arcsec$) is also detected, although its X-ray emission fades with time and becomes increasingly diffuse, appearing to merge with the upstream emission of knot A by 2023. At optical wavelengths, this feature has been interpreted as quasi-stationary feature \citepalias{Meyer2013}, but the X-ray emission is too faint to derive reliable proper-motion measurements.

        \subsubsection{The A-B-C complex} 
        \label{sec:A-B-C_complex}

            Knot A ($\sim12.4\arcsec$) is among the brightest X-ray features at all epochs, except during the HST-1 high state that remains visible in 2012. The feature exhibits negligible bulk motion in the deconvolved images, with $\beta_{\text{app}, \parallel} = (0.10 \pm 0.38)c$ and $\beta_{\text{app}, \perp} = (-0.17 \pm 0.38)c$, consistent with a stationary feature within uncertainties. The morphology of knot A also evolves with time, with the X-ray emission becoming more extended perpendicular to the jet axis in 2025. Optical studies resolve this knot into at least two subcomponents, A1 and A2, with apparent velocities of $\sim1.3c$ and $\sim0.3c$, respectively \citepalias{Meyer2013}. The observed X-ray morphological evolution may therefore reflect variations in the relative contributions of these subcomponents. In addition, a distinct feature is detected ($\sim0.5\arcsec$) downstream of knot A in 2017, which may correspond to the B1 component identified in optical observations. Combined with polarimetric studies indicating a compressed magnetic field oriented transverse to the jet axis (e.g., \citealt{Owen1989, Perlman1999, Harris2006, Avachat2016, Hada2024}), these results are consistent with the interpretation of knot A as a standing shock, possibly associated with an oblique reverse shock propagating upstream into the flow.
            
            Downstream, knot B marks a transition to a more dynamically complex regime. Optical observations resolve this region into multiple components (B1 to B5) and allow proper motion measurements for each, revealing a zig-zag morphology \citepalias{Meyer2013}. In our X-ray data, deconvolved images intermittently reveal substructures broadly consistent with this multi-component picture; however, the complexity of the region prevents reliable proper motion measurements for individual components. Nevertheless, several features can be tentatively cross-identified across wavelengths. The brightest X-ray emission is associated with B4 ($\sim14.3\arcsec$), while the downstream region corresponding to B5 ($\sim15.2\arcsec$) shows significant transverse evolution and appears fragmented in 2023. In addition, optical studies show two bar-like substructures connecting B3 and B4 (prominent around 2007--2008, between $\sim13\arcsec-14\arcsec$) exhibiting a central reduced-emission region ("hole"). The evolution of these structures, including transverse displacements, could naturally enhance or shift such gaps over time. Although B3 and B4 cannot be individually identified in X-rays, the observed transverse oscillations (see Fig.~\ref{fig:inter-epoch_frac-count-diff}) and the presence of a similar "hole" in 2023 (see Fig.~\ref{fig:subcomponents_hst-chandra}) suggest that a similar underlying multi-component structure is also present in X-rays, albeit only partially resolved.

            At knot C ($\sim17.5\arcsec$), the C-1 component, resolved into two substructures (c1 and c2) in optical observations \citepalias{Meyer2013}, exhibits significant variations in its X-ray apparent motion, with $\beta_{\text{app}, \parallel}$ changing from $(0.8 \pm 1.0)c$ to $(-1.1 \pm 0.4)c$ between epochs, likely reflecting shifts in the dominant subcomponent. The radial profile of C-1 is also more symmetric in X-rays than at infrared and optical wavelengths, where the emission exhibits a steeper downstream decline (see Fig.~\ref{fig:chandra-vs-hst-jwst}b). This behavior is consistent with the interpretation of knot C as a forward shock (\citealt{Harris2006}; \citetalias{Meyer2013}). The connection between C-2 and G-1 further leads to changes in the orientation of the combined structure, supporting the picture of a dynamically evolving shock region.

        \subsubsection{Downstream of knot C}
        \label{sec:downstream_knot-C}
            Beyond knot C, the X-ray emission becomes increasingly misaligned with structures seen at longer wavelengths, although counterparts can still be identified. While the emission at optical and infrared wavelengths becomes fainter relative to knot C-1 in this extended region, the X-ray brightness remains relatively enhanced, as indicated by the measured ratios between the newly detected feature (region labeled "Extended" in Fig.~\ref{fig:psf-deconv-emc2_epochs}) and C-1 (0.41 in the 2025 X-ray data versus 0.10 at 3.56~$\mu$m with JWST). Although the signal remains limited and does not allow robust proper motion or flux variability measurements, the X-ray morphology of the new feature suggests ongoing evolution and possible brightening, as it was not detectable in 2012--2017. These properties are consistent with a localized acceleration site, potentially linked to jet-environment interaction (e.g., with an obstacle). 

        \subsubsection{Resolution and component blending effects} 
        \label{sec:resolution_blending_effects}
        
            The comparison between raw and deconvolved images shows that limited angular resolution can significantly bias the inferred kinematics, particularly in regions where multiple components are blended. In such cases, the measured centroid traces a flux-weighted average of unresolved substructures, causing the inferred motion to vary as the relative brightness of the components evolves. This effect primarily impacts the parallel proper motions, where multiple components moving along the jet axis are more strongly blended. This is clearly illustrated in HST-1, where apparent velocities range from $\sim2c$ to $\sim6c$, despite centroid shifts of only $\sim0.08\arcsec$ over a 5~yr baseline. A similar effect is observed at knot D: X-ray measurements report $\beta_{\text{app}, \parallel} = (2.4 \pm 0.6)c$ \citepalias{Snios2019}, consistent with our raw-image measurements of $(2.1 \pm 0.4)c$, but lower velocities are obtained in later epochs ($\sim1c$). This discrepancy reflects the coexistence of multiple components with distinct velocities, including D-E ($\sim0.3c$), D-M ($\sim4.3c$), and D-W ($\sim1c$) \citepalias{Meyer2013}, whose varying flux contributions can shift the measured centroid. These results highlight the importance of high-resolution imaging and long-term monitoring for disentangling the multi-component structure of the jet and correctly interpreting its kinematic and evolution.

    \subsection{Structures around the nucleus}
        In addition to the collimated jet and knots, the PSF-deconvolved images reveal faint asymmetric structures within $\lesssim1.5\arcsec$ ($\sim120$~pc) of the AGN. These features are present across all epochs and display noticeable morphological variations, ranging from compact clumpy features to broader extended emission. Their reconstructed count levels are comparable to those of the fainter, more diffuse regions of the jet, although their reconstruction stability is relatively low ($\sim 0.3-0.4~\sigma$~pix$^{-1}$; see Section~\ref{sec:psf-deconvolved_images}) and approaches the background level; they should therefore be interpreted with caution.

        The observed morphology may indicate the presence of a circumnuclear component, possibly associated with the soft and multiphase gas previously detected within the central $\sim100$~pc, including material influenced by interactions with the jet and/or a broader nuclear outflow (e.g., \citealt{Sparks2004, Russell2018, Osorno2023, Ray2024}). The observed spatial scales are also comparable to commonly adopted estimates of the Bondi radius in M87 ($1.5\arcsec-2.8\arcsec$; e.g., \citealt{Bondi1952, Russell2015}), suggesting that these structures could trace the transition region where gas supply and AGN feedback interact. Deeper observations and additional multi-epoch imaging will be required to confirm their reality and clarify their physical nature.

\section{Conclusions}
\label{sec:conclusions}

    We have presented a 13-year (2012--2025) \textit{Chandra} HRC-I study of the M87 jet, using PSF deconvolution to obtain the highest-resolution X-ray view of the jet to date. The reconstructed images reveal previously blended substructures throughout the jet and enable us to track their morphological and kinematic evolution across four epochs. Our principal results are as follows~:

        \begin{enumerate}
            \item \textit{Global flux decrease and synchrotron cooling.} Flux measurements performed on the raw images reveal a decrease in X-ray emission across much of the jet between 2012 and 2025, reaching up to $84\%$ in HST-1 and $27\%$ in knot A. This behavior is consistent with synchrotron cooling of the highest-energy electrons. Modeling the observed fading yields minimum magnetic field strengths of $\sim 324$--$1006~\mu$G for HST-1 and $\sim 41$--$115~\mu$G for knot A, broadly consistent with independent equipartition and variability-based estimates. 
            
            \item \textit{Deconvolved substructure.} The combination of HRC imaging, time-dependent empirical PSFs, and EMC$^2$ deconvolution reveals complex and evolving substructure throughout the jet. In particular, HST-1 is resolved into two distinct components from 2023 onward, closely matched to counterparts in infrared (JWST) and optical (HST) images. These structures are heavily blurred in the raw images and demonstrate the complex multi-component morphology of the X-ray jet on sub-arcsecond scales.

            \item \textit{Multi-wavelength morphology.} Comparison with contemporaneous millimeter (ALMA), infrared (JWST), and optical (HST) observations shows that the principal X-ray knots are co-spatial with their counterparts, while the X-ray emission peaks are generally offset upstream by $\sim0.07\arcsec-0.5\arcsec$. These offsets are consistent with the expectation from synchrotron cooling, wherein the shortest-lived, highest-energy electrons trace the most recent acceleration sites.

            \item \textit{Jet dynamics and proper motions.} The jet exhibits a rich variety of dynamical behavior, including superluminal motions, transverse displacements, evolving inter-knot emission, and the emergence of new subcomponents. Apparent velocities of up to $4.8~c$ are observed in HST-1, while a few regions remain quasi-stationary, consistent with standing shocks and recollimation features. The deconvolved images further demonstrate that unresolved blending of nearby components with different velocities can strongly bias inferred proper motions and morphology, emphasizing the importance of high angular resolution for reliable kinematic studies of relativistic jets. The coexistence of stationary and moving structures, together with the disturbed morphology observed throughout the flow, points to a dynamically complex jet shaped by shocks, instabilities, and broader MHD processes.
        \end{enumerate}
    Beyond the scientific results, this work demonstrates the capability of \textit{Chandra}/HRC-I to resolve and monitor sub-arcsecond structures in nearby X-ray jets when combined with modern image reconstruction techniques. The combination of high-quality HRC-I data, precise astrometric alignment, time-dependent empirical PSFs, and EMC$^2$ deconvolution enables a more faithful reconstruction of the intrinsic jet morphology than previously possible. This approach opens a new window on the long-term evolution of relativistic jets and highlights the continuing unique capabilities of \textit{Chandra} for studying particle acceleration and jet dynamics on scales inaccessible to other X-ray observatories. Further pushing the limits of high-resolution X-ray imaging through deeper observations and more closely spaced monitoring epochs will improve sensitivity to faint inter-knot and transverse structures, better constrain their evolution, and provide deeper insight into the physics governing particle acceleration and energy dissipation in relativistic jets.

\begin{acknowledgments}
\small
    We thank Vinay Kashyap for generating the empirical PSFs used in this work and for valuable discussions. 
    CP acknowledges support from the Fonds de recherche du Québec through awards 346450 and 2003876. Support for this work was provided by the National Aeronautics and Space Administration through Chandra Award Number GO1-22095X issued by the Chandra X-ray Center, which is operated by the Smithsonian Astrophysical Observatory for and on behalf of the National Aeronautics Space Administration under contract NAS8-03060. G.S. and R.K. acknowledge support from the Smithsonian Institution and the Chandra High Resolution Camera Project through NASA contract NAS8-03060. MLGM acknowledges financial support from NSERC via the Discovery grant program and the Canada Research Chair program (110-2024-2025-Q1-4328, 207-2024-2025-Q1-01314, 110-2024-2025-Q1-4327).
    This research has made use of data obtained from the \textit{Chandra} Data Archive provided by the \textit{Chandra} X-ray Center (CXC). 
    This research is based on observations made with the NASA/ESA Hubble Space Telescope obtained from the Space Telescope Science Institute, which is operated by the Association of Universities for Research in Astronomy, Inc., under NASA contract NAS 5–26555. These observations are associated with programs 12989, 14618, and 17592.
    This work is based on observations made with the NASA/ESA/CSA James Webb Space Telescope. The data were obtained from the Mikulski Archive for Space Telescopes at the Space Telescope Science Institute, which is operated by the Association of Universities for Research in Astronomy, Inc., under NASA contract NAS 5-03127 for JWST. These observations are associated with JWST program GO-3055.
    All the HST and JWST data used in this paper can be found in MAST: \dataset[10.17909/h624-0v53]{https://doi.org/10.17909/h624-0v53.}. 
    This paper makes use of the following ALMA data: ADS/JAO.ALMA\#2016.1.01154.V. ALMA is a partnership of ESO (representing its member states), NSF (USA) and NINS (Japan), together with NRC (Canada), NSTC and ASIAA (Taiwan), and KASI (Republic of Korea), in cooperation with the Republic of Chile. The Joint ALMA Observatory is operated by ESO, AUI/NRAO and NAOJ. The National Radio Astronomy Observatory and Green Bank Observatory are facilities of the U.S. National Science Foundation operated under cooperative agreement by Associated Universities, Inc.
\end{acknowledgments}

\bibliography{references}{}
\bibliographystyle{aasjournalv7}

\appendix 
\section{Flux evolution of HST-1 and knot A}
\label{appendix:knot_flux-evolution}
\restartappendixnumbering

    \begin{figure*}[h]
        \centering
        \includegraphics[width=0.5\textwidth]{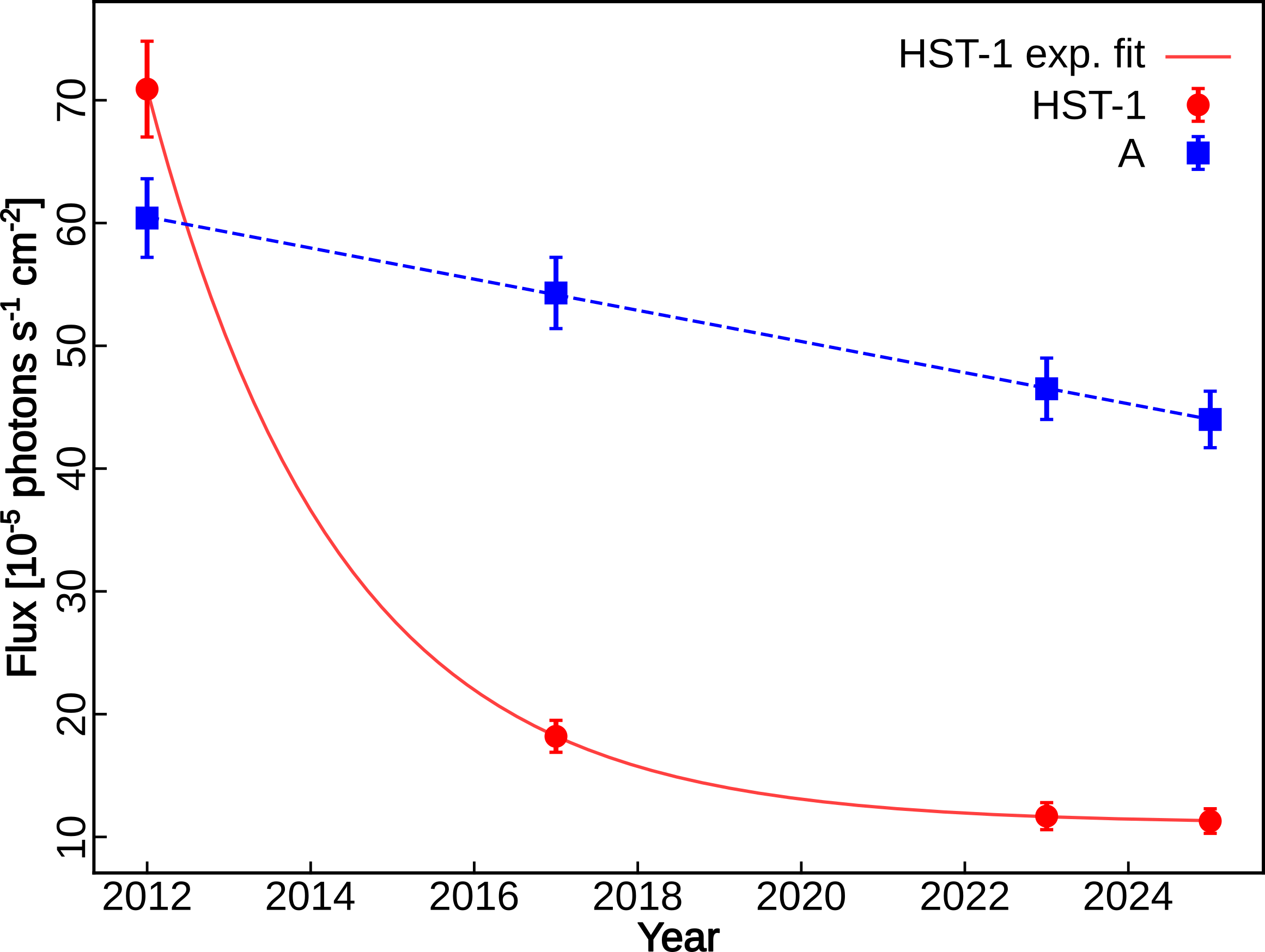}
        \caption{Flux evolution of HST-1 (red circles) and knot A (blue squares) as a function of time. Knot A exhibits a linear decline over 12.9~yr, whereas HST-1 follows an exponential decay. The solid red curve shows the best-fit exponential model to HST-1, yielding an observed decay timescale of $(2.3 \pm 0.3)$~yr.\label{fig:flux-evolution_HST1-A}}
    \end{figure*}

\section{Inter-epoch fractional difference maps}
\label{appendix:frac-count-diff}
\restartappendixnumbering

     \begin{figure*}[h]
        \centering
        \includegraphics[width=\textwidth]{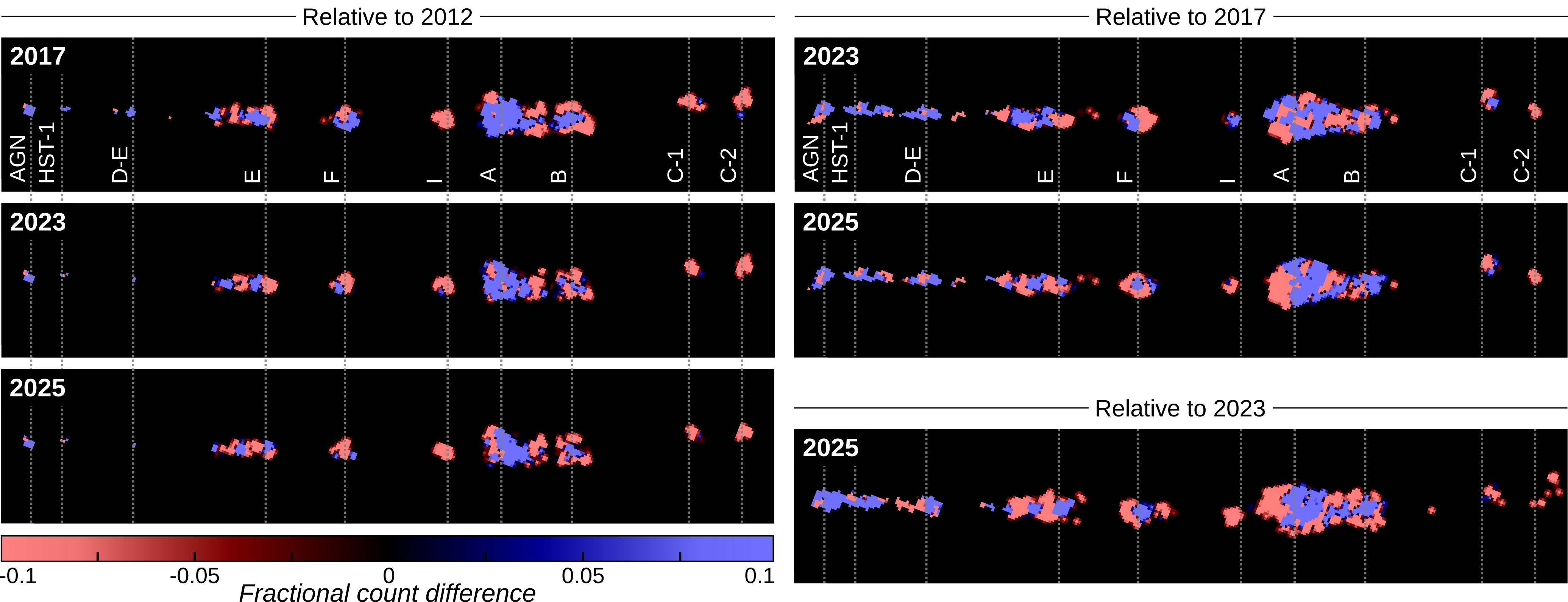}
        \caption{\footnotesize Fractional count (PSF-deconvolved, $0.0659\arcsec$~pix$^{-1}$) difference maps relative to the reference epoch indicated, defined as $(C_2 - C_1)/C_1$.  The maps highlight spatial and morphological variations independent of absolute flux differences, with blue/red regions indicating flux increases/decreases. For each panel, the AGN and knots are marked with dotted vertical lines. All maps are smoothed with a Gaussian kernel ($\sigma = 2$~pix).\label{fig:inter-epoch_frac-count-diff}}
    \end{figure*}

\section{Proper motion measurements}
\label{appendix:proper_motions}
\restartappendixnumbering

    \begin{figure*}[h]
        \centering
        \includegraphics[width=0.75\textwidth]{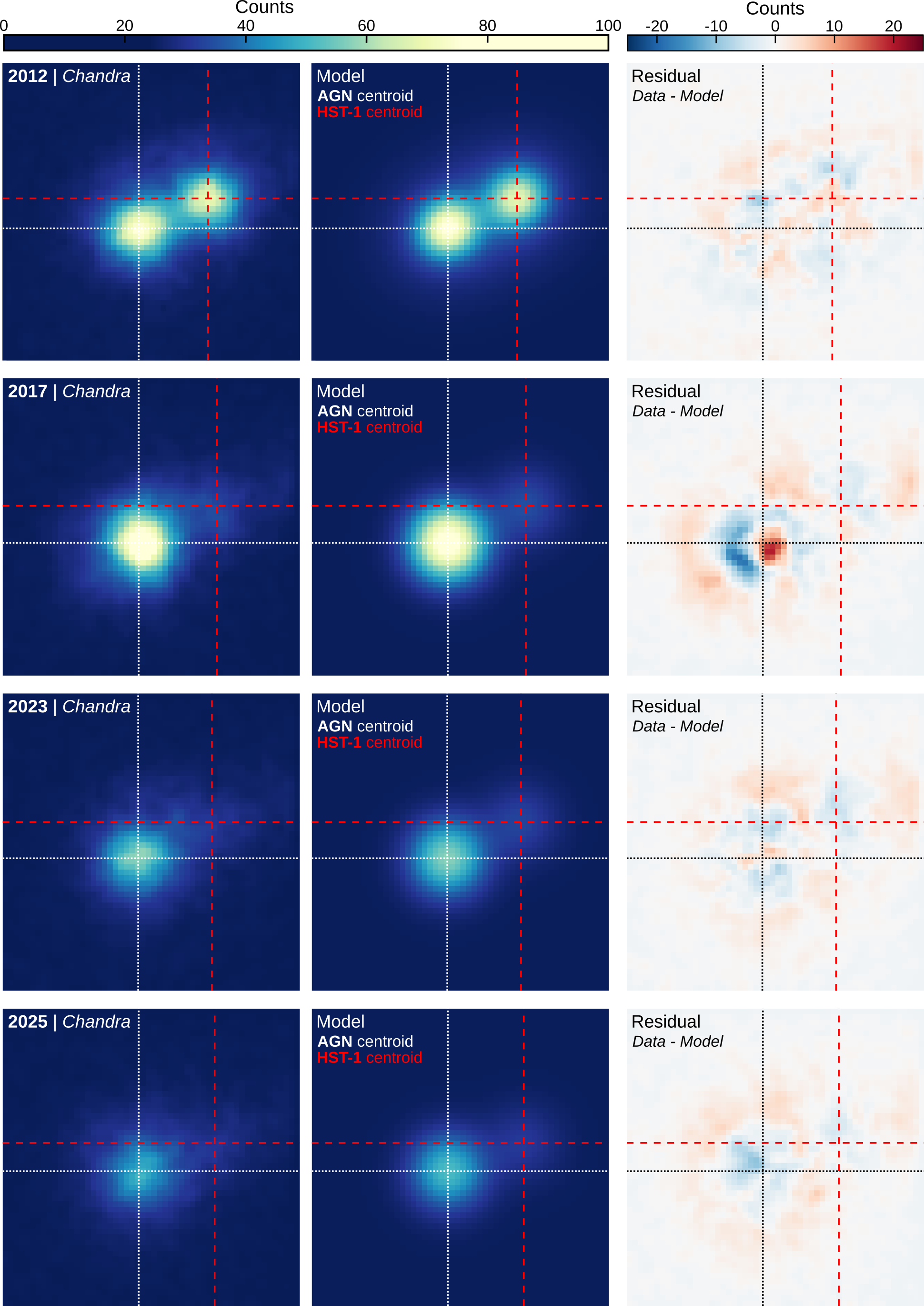}
        \caption{\footnotesize \textit{Chandra} count images (\textit{left}; binned to a pixel size of $0.0659\arcsec$), corresponding best-fit 2D models obtained with \texttt{Sherpa} (\textit{center}), and residual maps (\textit{right}; defined as data $-$ model) from 2012 to 2025 (top to bottom). The dotted white/black and dashed red lines mark the fitted centroid positions of the AGN and HST-1, respectively.\label{fig:chandra-model_agn-hst1}}
    \end{figure*} 

    \begin{deluxetable*}{lcccccccccccccc}
    \tablecaption{Distances and proper motions measured from raw (non-deconvolved) images. Raw-image proper motions may be biased by unresolved component blending (see Section~\ref{sec:resolution_blending_effects}).\label{tab:proper_motions_raw}}
    \tabletypesize{\scriptsize}
    \tablewidth{0.94\textwidth}
    \tablewidth{0pt}
    \tablecolumns{14}
        \tablehead{\colhead{Knot} & \colhead{Dist. [$\arcsec$]\tablenotemark{a}} & \multicolumn{3}{c}{$\mu_\parallel$ [mas yr$^{-1}$]} & \multicolumn{3}{c}{$\mu_\perp$ [mas yr$^{-1}$]} & \multicolumn{3}{c}{$\beta_{\mathrm{app},\parallel}$ [$c$]} & \multicolumn{3}{c}{$\beta_{\mathrm{app},\perp}$ [$c$]} \\
        \cline{3-5}\cline{6-8}\cline{9-11}\cline{12-14}
        \colhead{} & \colhead{} & \colhead{2017} & \colhead{2023} & \colhead{2025} & \colhead{2017} & \colhead{2023} & \colhead{2025} & \colhead{2017} & \colhead{2023} & \colhead{2025} & \colhead{2017} & \colhead{2023} & \colhead{2025}}
        \startdata
        HST-1 & 0.92
              & 26.2(2.8) &  6.8(3.7) &  5.1(1.3)
              &  8.5(2.3) &  4.6(3.0) & -3.7(1.1)
              &  6.9(0.6) &  1.8(0.9) &  1.4(0.3)
              &  2.2(0.5) &  1.2(0.7) & -1.0(0.3) \\
        D     & 2.69
              &  8.1(1.7) &  3.7(0.8) &  4.1(0.8)
              &  0.6(1.5) & -1.1(0.7) &  0.4(0.7)
              &  2.1(0.4) &  1.0(0.2) &  1.1(0.2)
              &  0.2(0.4) & -0.3(0.2) &  0.1(0.2) \\
        D-W   & 4.43
              &  9.6(4.2) &  4.7(1.9) &  1.9(1.7)
              & -0.7(2.9) & -1.0(1.3) & -2.4(1.2)
              &  2.5(1.1) &  1.2(0.5) &  0.5(0.4)
              & -0.2(0.8) & -0.3(0.4) & -0.6(0.3) \\
        E     & 5.90
              & -0.7(3.5) & -4.4(1.6) & -4.4(1.4)
              & -3.1(2.8) &  0.0(1.3) &  0.0(1.2)
              & -0.2(0.9) & -1.2(0.4) & -1.2(0.4)
              & -0.8(0.7) &  0.0(0.3) &  0.0(0.3) \\
        F     & 8.83
              &  9.4(3.6) &  3.6(1.7) &  3.5(1.5)
              & -4.0(3.3) & -0.9(1.5) &  0.3(1.3)
              &  2.5(0.9) &  0.9(0.5) &  0.9(0.4)
              & -1.1(0.9) & -0.2(0.4) &  0.1(0.4) \\
        A     & 12.41
              &  7.7(1.5) &  5.5(0.8) &  4.0(0.7)
              & -3.9(1.5) & -1.5(0.7) &  1.9(0.7)
              &  2.0(0.4) &  1.5(0.2) &  1.0(0.2)
              & -1.0(0.4) & -0.4(0.2) &  0.5(0.2) \\
        B     & 14.44
              &  0.9(3.6) &  1.5(1.6) & -0.4(1.5) 
              & -1.9(3.2) &  0.6(1.5) &  1.8(1.3)
              &  0.2(1.0) &  0.4(0.4) & -0.1(0.4)
              & -0.5(0.9) &  0.2(0.4) &  0.5(0.3) \\
        C     & 17.48
              &  7.6(5.0) &  5.1(2.3) &  0.7(2.0)
              &  5.4(4.9) &  0.9(2.3) & -0.2(2.0)
              &  2.0(1.3) &  1.3(0.6) &  0.2(0.5)
              &  1.4(1.3) &  0.2(0.6) & -0.1(0.5) \\
        \enddata
        \tablenotetext{a}{Projected distance from the core measured in the 2012 reference image.}
        \tablenotemark{\textbf{Notes.} Negative apparent motions likely result from unresolved substructures and their evolving flux contributions.}
    \end{deluxetable*}

\end{document}